\documentclass[trackchanges]{aastex62}

\usepackage{multirow}
\usepackage{rotating}
\usepackage{epstopdf}
\usepackage{color}
\newcommand{\tzl}{\textbf}

\shorttitle{Superflares on solar-type stars from TESS}
\shortauthors{Tu et al.}

\begin{document}

\title{Superflares on solar-type stars from the first year observation of {\em TESS}}


\correspondingauthor{F. Y. Wang}
\email{fayinwang@nju.edu.cn}

\author[0000-0001-6606-4347]{Zuo-Lin Tu}
\affil{School of Astronomy and Space Science, Nanjing University, Nanjing 210093, China}

\author[0000-0002-6926-2872]{Ming Yang}
\affil{School of Astronomy and Space Science, Nanjing University, Nanjing 210093, China}
\affil{Key Laboratory of Modern Astronomy and Astrophysics (Nanjing University), Ministry of Education, Nanjing 210093, China}

\author{Z. J. Zhang}
\affil{School of Astronomy and Space Science, Nanjing University, Nanjing 210093, China}

\author[0000-0003-4157-7714]{F. Y. Wang}
\affil{School of Astronomy and Space Science, Nanjing University, Nanjing 210093, China}
\affil{Key Laboratory of Modern Astronomy and Astrophysics (Nanjing University), Ministry of Education, Nanjing 210093, China}



\begin{abstract}
Superflares, which are strong explosions on stars, have been well studied
with the progress of space time-domain astronomy. In this work, we
present the study of superflares on solar-type stars using
{\em Transiting Exoplanet Survey Satellite} ({\em{TESS}}) data. Thirteen sectors
of observations during the first year of the {\em TESS} mission have
covered the southern hemisphere of the sky, containing 25,734
solar-type stars. We verified 1216 superflares on 400 solar-type
stars through automatic search and visual inspection with 2 minute
cadence data. Our result suggests a higher superflare frequency
distribution than the result from {\em Kepler}. This may be because the majority of {\em TESS} solar-type stars in our dataset are
rapidly rotating stars. The power-law index $\gamma$ of the
superflare frequency distribution ($dN/dE\propto E^{-\gamma}$) is
constrained to be $\gamma = 2.16\pm 0.10$, which is a little larger
than that of solar flares but consistent with the results from {\em
Kepler}. Because only seven superflares of Sun-like stars are detected,
we cannot give a robust superflare occurrence frequency. Four
stars are accompanied by unconfirmed hot planet candidates.
Therefore, superflares may possibly be caused by stellar magnetic
activities instead of planet-star interactions. We also find an
extraordinary star, TIC43472154, which exhibits about 200 superflares
per year. In addition, the correlation between the energy and duration
of superflares ($T_{\text {duration }} \propto E^{\beta}$) is
analyzed. We derive the power-law index to be $\beta=0.42\pm0.01$,
which is a little larger than $\beta=1/3$ from the prediction
according to magnetic reconnection theory.

\end{abstract}

\keywords{stars: flare - stars: solar-type}


\section{Introduction}
\label{sec:intro}

Solar activities are closely connected with the lives of human
beings. For example, solar winds can affect space weather, and
trigger the geomagnetic storms on the Earth. Solar activities thus
have large impacts on many fields, e.g. spacecraft electronics,
commercial aviation, and radio communication
\citep{2011SpWea...9.6001C}. The Carrington event
\citep{1859MNRAS..20...13C}, which generated a very large solar
flare with a total energy up to $10^{32}$ erg, caused widespread
disruption of telegraph systems and is considered to be one of the
most severe solar storms to date. An indication of this event is also found in
polar ice \citep{2006AdSpR..38..232S}. However, the reliability of
nitrate in ice records as a proxy for solar flares is strongly
debated \citep{Melott16,Mekhaldi17}.

Solar flares are also studied by indirect methods, such as rapid
increases of $^{14}$C in tree rings. Four such events have been
found \citep{Miyake12,Miyake13,2017NatCo...8.1487W,Park17}. Based on the fact that
quasi-simultaneous peaks of $^{10}$Be and $^{36}$Cl are found to be
associated with the peaks of $^{14}$C, solar superflares are considered as
their physical origin \citep{Mekhaldi15,Miyake19}.
However, the peaks of $^{10}$Be and $^{36}$Cl are adjusted to fit
the $^{14}$C peaks \citep{Mekhaldi15}. So, if the peaks really are
simultaneous, a solar origin is the most probable
\citep{Mekhaldi15,Miyake19}. If these peaks are not correlated,
other models may be possible
\citep{Neuhauser14,Neuhauser15,Wang19}.

Superflares are much stronger explosions than typical solar flares,
with total energies varying from $10^{33}$ -- $10^{38}\;{\rm erg}$ and
durations of longer than an hour. The energy of a superflare is so high
that it can directly harm nearby living
creatures. This has therefore attracted people's interest as to whether our
Sun can generate superflares, just like the nine main sequence F8--G8
stars near our solar system \citep{2000ApJ...529.1026S}. Previous
studies have revealed a power-law relation between flare energy and
frequency $dN/dE\propto E^{-\gamma}$ \citep{1985SoPh..100..465D},
which can be explained by self-organized criticality happening in a
nonlinear energy dissipation system
\citep[e.g.][]{Bak87,Lu91,Aschwanden11,2013NatPh...9..465W}. For
hard X-ray solar flares, $\gamma$ was estimated to be $1.53\pm 0.02$
by \citet{1993SoPh..143..275C}. For nanoflares and microflares,
$\gamma$ is $1.79\pm 0.08$ \citep{2000ApJ...535.1047A} and 1.74
\citep{1995PASJ...47..251S}, respectively. With the help of {\em
Kepler} space telescope, \citet{2012Natur.485..478M} detected 101
superflares on 24 slowly rotating solar-type stars and derived
$\gamma = 2.0 \pm 0.2$, which is very similar to that of the
distribution of solar flares. This therefore suggests there is a possibility
for superflares to occur on the Sun, but with a comparatively lower
frequency than rapidly rotating G-type stars.

There have been many observational and theoretical studies since the
discovery of superflares on G-type stars with {\em Kepler} data.
\citet{2013ApJS..209....5S} and \citet{2013ApJ...771..127N} studied
the connection between superflares and starspots by using more
samples, and concluded that the energy of superflares is related to
the starspot coverage, which is similar to the relation between
solar flares and sunspots. Note that although the samples from
\citet{2013ApJS..209....5S} were mistakenly mixed with some stars
which are not in the main sequence, their conclusions are not
changed \citep{2019ApJ...876...58N}. Large starspots can be
generated by solar dynamo mechanism \citep{2013PASJ...65...49S} and
may store energy for superflares. Superflares, therefore, may have
the same origin as solar flares. Because the {\em Kepler} input
catalog may not give an accurate estimation of the properties of solar-type
stars, ground-based follow-up observations were also dedicated to
the research, which has constrained stellar periodicity and spectrum
information \citep[][]{2015aN,2015bN}. \citet{2015EP&S...67...59M}
supported the comparability by extending the flares to lower
energies using the 1 minute short-cadence data from {\em Kepler}.
\citet{2016NatCo...711058K} also provided support by analyzing the
chromospheric activities of 5648 Sun-like stars and 48 superflare
stars in the field of view of both {\em Kepler} and the Large Sky Area
Multi-Object Fibre Spectroscopic Telescope (LAMOST).

On the other hand, there are different explanations for superflares.
\citet{2000ApJ...529.1031R} argued that a hot Jupiter companion can
cause superflares on a solar-type star. Numerical simulations by
\citet{2004ApJ...602L..53I} also presented the importance of
the effect of planet-star magnetic interactions on stellar activities. In
addition, \citet{2015ApJS..221...18H} argued that the rotational
modulation may be caused by faculae instead of starspots. It is also
possible that flares and the magnetic features that dominate
rotational modulation may possibly have different source regions
\citep{2018ApJS..236....7H}.

In this paper, we aim to detect superflares on solar-type stars
using {\em Transiting Exoplanet Survey Satellite} ({\em TESS}) data. The
primary goal of  {\em TESS} is to find planets around bright nearby
stars to enable further ground-based follow-up observations
\citep{2015JATIS...1a4003R}. {\em TESS} has three advantages for
studying stellar flares. Firstly, the stars observed by {\em TESS}
are bright enough to achieve a high signal-to-noise ratio. Secondly,
the 2 minute cadence allows the study of more detailed flare
properties, such as duration and energy. Besides, to obtain more
credible stellar parameters, the {\em TESS} input catalog (TIC) has
imported data from other space and ground-based projects, such as
{\em Gaia}-DR1 and 2 and LAMOST-DR1 and DR3.

This paper is constructed as follows. In Section \ref{sec:DM}, we
describe the methods to select solar-type stars and superflare
candidates from the {\em TESS} data. Stellar periodicity and flare
energy are also derived. The main results are described and
discussed in Section \ref{sec:results}, including the occurrence
frequency of superflares, active flare stars, systems with
exoplanets, and the correlation between the duration and energy of
superflares. Finally, we summarize our findings in Section
\ref{sec:conclusions}.

\section{Data and methods}\label{sec:DM}

\subsection{Selection of Solar-type stars}\label{sec:solar-type data}
{\em TESS} was launched on 2018 April 18, and carries four
identical cameras. During its first year of observations, {\em TESS}
has scanned the southern hemisphere of the sky and obtained data
products for 13 segments (sector 1 - sector 13). Each segment covers
about 27 days. In this work, we adopt the presearch data conditioned
(PDC) light curves to avoid the instrumental systematics.

Firstly, the selection criteria of solar-type stars and
Sun-like stars should be clarified. Solar-type stars are selected
according to following criteria: (1) the surface effective
temperature satisfies $5100 \mathrm{K} \leqslant
T_{\mathrm{eff}}<6000 \mathrm{K}$ and (2) the surface gravity in
$\log$ scale is $\log g>4.0$
\citep{2000ApJ...529.1026S,2012Natur.485..478M}. Those
solar-type stars with \tzl{$5600 \mathrm{K} \leqslant
T_{\mathrm{eff}}<6000 \mathrm{K}$} and stellar periodicity
$>10$ days are considered as Sun-like stars. In total, 26,034
targets meet the requirements based on the latest TIC v8 \citep[][]{2019AJ....158..138S}, which is expected
to be the last version of the TIC. Then, we examine these
solar-type stars with the {\em Hipparocos}--2 catalog
\citep{2007A&A...474..653V} to exclude some confirmed binary stars.
In total, 145 stars are excluded from the dataset.

In contrast to {\em Kepler}, of which the pixel scale is about 4$ ''$, {\em TESS} has a larger pixel scale of 21$ ''$. 90\% of the 
energy are ensquared by 4$\times$4 pixels, in other words, it is encircled by a
radius of 42$ ''$ \citep{2015JATIS...1a4003R}. Therefore, it is possible
that in one pixel from {\em TESS}, the primary object is contaminated
by other stars. Therefore, we use {\em Gaia}-DR2
\citep{2018A&A...616A...1G} to search for stars within a 42$ ''$ radius
near the primary solar-type star. Within a 21$ ''$ radius, we find
155 stars containing other brighter stars, which are excluded from
the dataset. The reasons for this are: (1) flux from these brighter stars may
significantly affect the light curves of the main target; and (2) in just
one pixel scale (21$ ''$), we cannot separate these brighter
stars apart from main targets. Next, 2849 solar-type stars, of
which the nearby stars show surface effective temperature within
3000K--4000K, are flagged as possessing M dwarf candidates. The
reason why they are not excluded from the dataset is shown in
Appendix \ref{sec:KvsT}.

\subsection{Selection of superflares}
\label{sec:superflares-selection}

The PDC light curves of the solar-type stars are used to search for
superflares. There are several selection methods in the
literature. For example, clean light curves with only flare
candidates can be acquired by fitting the quiescent variability
\citep[e.g.][]{2011AJ....141...50W,2015ApJ...798...92W,2019ApJS..241...29Y}.
In addition, flare candidates can also be obtained by calculating
the distribution of brightness changes between consecutive data
points
\citep[e.g.][]{2012Natur.485..478M,2013ApJS..209....5S,2015EP&S...67...59M}.
In this paper, we choose the latter method to detect superflares. According
to \citet{2015EP&S...67...59M}, brightness changes between two pairs
of consecutive points $\Delta F^{2}$ are defined as :
\begin{equation}\label{equ:deltaF}
\Delta F^{2}\left(t_{i,
n}\right)=s\left(F_{i}-F_{i-n-1}\right)\left(F_{i+1}-F_{i-n}\right),
\end{equation}
where $F_i$ and $t_i$ represent the flux and the time of the $i$th
data point, $n$ is an integer number, and $s=\pm 1$. $s= 1$ when
both $\left(F_{i}-F_{i-n-1}\right) >0$ and
$\left(F_{i+1}-F_{i-n}\right)>0$, otherwise $s=-1$. $\Delta F^{2}$
during a superflare event will become much larger than the quiescent
situation. We set $n=2$ to detect flare candidates with rise times
larger than 4 minutes \citep{2015EP&S...67...59M}. Besides, we also
set $n=5$ and obtained another group of $\Delta F^{2}$ in case of
missing flares with longer rise times. A data point is recognized as
a flare candidate when its $\Delta F^{2}$ is at least three times larger than the value
at the top $1\%$ of the $\Delta F^{2}$ distribution. We then located the peak data point for each flare candidate.
To obtain a complete flare event, we use the data from 0.05 to 0.01
days before the peak and from 0.05 to 0.10 days after the peak in
the case of $n=2$, and use the data from 0.15 to 0.03 days before
the peak and from 0.15 to 0.25 days after the peak  in the case of
$n=5$. A quadratic function, $F_{q}(t)$, is adopted to remove the
long-term stellar variability. The start time and end time of each
flare candidate are the first and last point when $F(t)-F_{q}(t)$ is
three times larger than the photometric error. We only reserve a
flare candidate when (1) there are at least three consecutive points
during the spike event, and (2) the decay time is longer than the
rise time.

\begin{figure*}[!hp]
\centering
\includegraphics[width=1\linewidth]{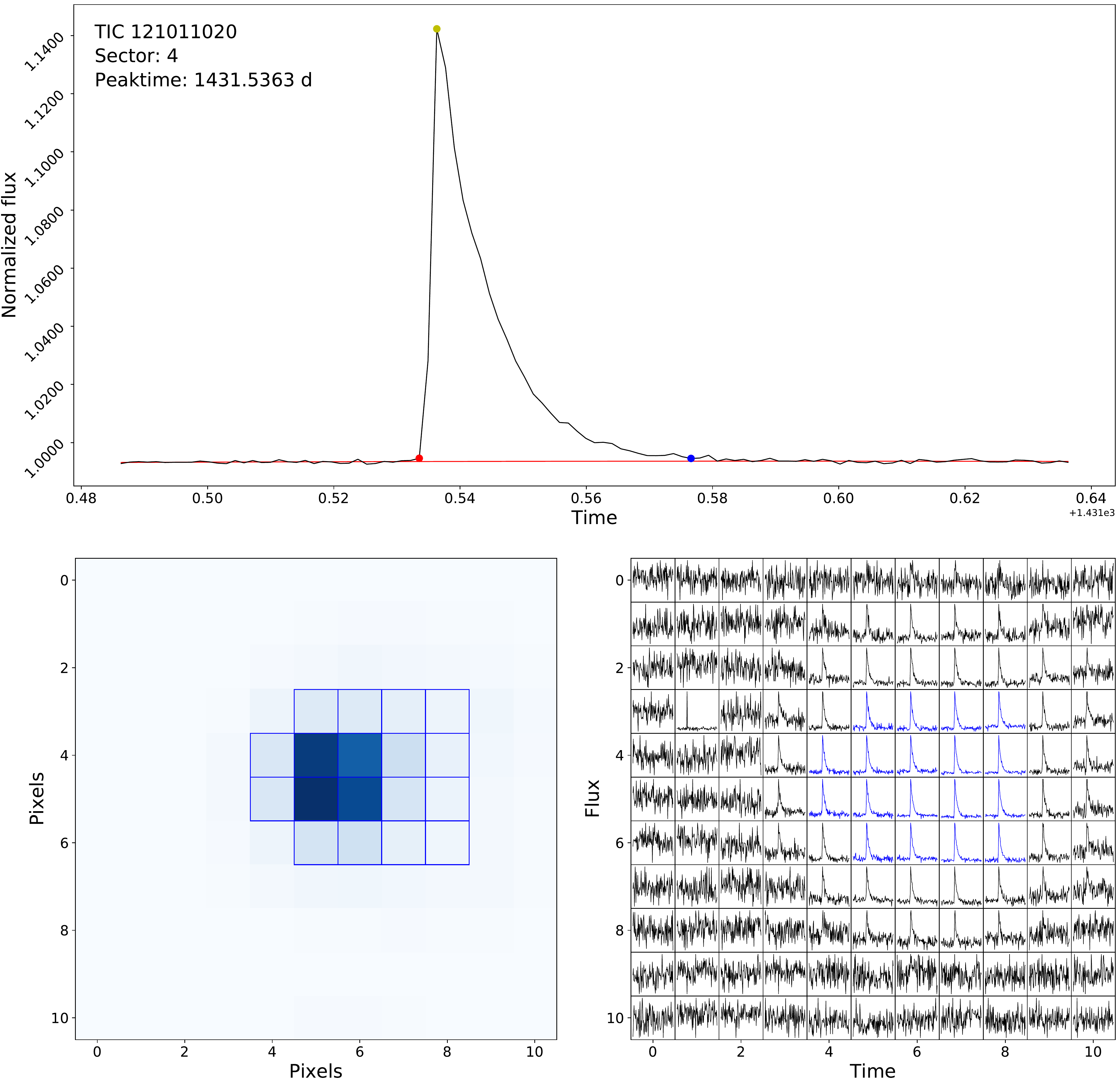}
\caption{An example of a true flare event. The upper panel gives a flare light curve of TIC121011020. The black solid line stands for
normalized flux $F(t)$. The flux fitted by the quadratic function ($F_{\rm
q}{(t)}$) is shown as red solid line. The red, yellow, and blue
points represent the beginning, peak, and end times of the flare,
respectively. The light curve in this panel shows a standard flare
shape with a rapid rise and slow decay. The lower-left panel shows
pixel-level data at the peak time. The blue frames encircle {\em
TESS} pipeline aperture masks. Those pixels are masked with a  deeper
blue where the flux is greater than the others. In the lower-right
panel, we present the light curves of each target pixel, which
corresponds to the every single block in the lower-left panel. The blue
light curves stand for {\em TESS} pipeline aperture masks. We may
conclude from this panel that the flare-shape light curves are
distributed as a point-spread function (PSF). }
        \label{fig:flare_sample_a}
    \end{figure*}

\begin{figure*}[!h]
\centering
\includegraphics[width=1\linewidth]{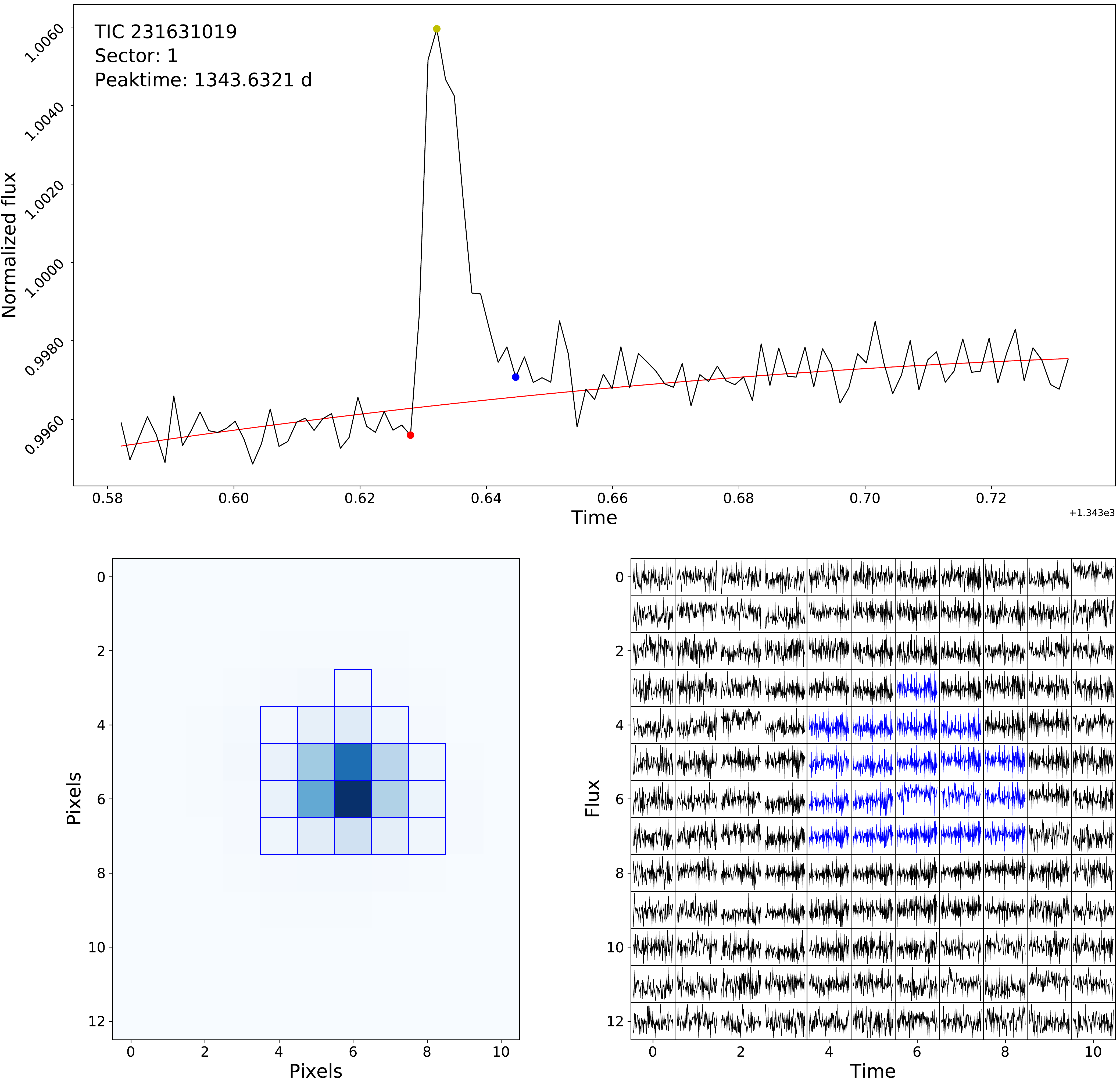}
\caption{Same as Figure \ref{fig:flare_sample_a}, but for a false
event or a very weak flare, which is excluded from the flare candidates. In the lower-right panel, unlike what we perceive from Figure
\ref{fig:flare_sample_a}, it may not show light curves distributed
as a PSF. But the noise level flux has been flooded in all pixels.}
\label{fig:flare_sample_b}
\end{figure*}

Then, we check each flare candidate by using pixel-level data to
exclude false positives such as eclipses, random flux jumps, and
cosmic rays. Firstly, we remove the candidates occurring
simultaneously on different stars. Second, the pixel-level light
curves during the flare event should match the optimal aperture
given by {\em TESS}. Thirdly, visual inspection is applied to ensure
the flare-like shape and obvious flux increases in pixel-level light
curves as suggested by \citet{2015ApJ...798...92W}. We present the
example of a true flare event in Figure \ref{fig:flare_sample_a}.
The light curve of TIC121011020 in sector 4 perfectly shows a
standard case of a flare-shaped curve, with rapid rise and slow decay.
Figure \ref{fig:flare_sample_b} shows a case with pixel-level
fluctuations under the noise level. We exclude it from superflares
since it is more likely to be a very weak flare or just noise.
After excluding all the false positives, we finally find 1216
superflares occurring on 400 solar-type stars.

\subsection{Periodicity estimation}
\label{sec:periodic_estimation} 
The periodicity of the solar-type stars
are estimated through the Lomb--Scargle method
\citep{1976Ap&SS..39..447L,1982ApJ...263..835S}, which is suitable
for unevenly sampled data \citep{2018ApJS..236...16V}. We set the
false alarm probability (FAP) to $10^{-4}$ to search the stellar
period \citep[e.g.][]{2019arXiv190900189C}.

\begin{figure*}
\centering
\includegraphics[width=0.6\linewidth]{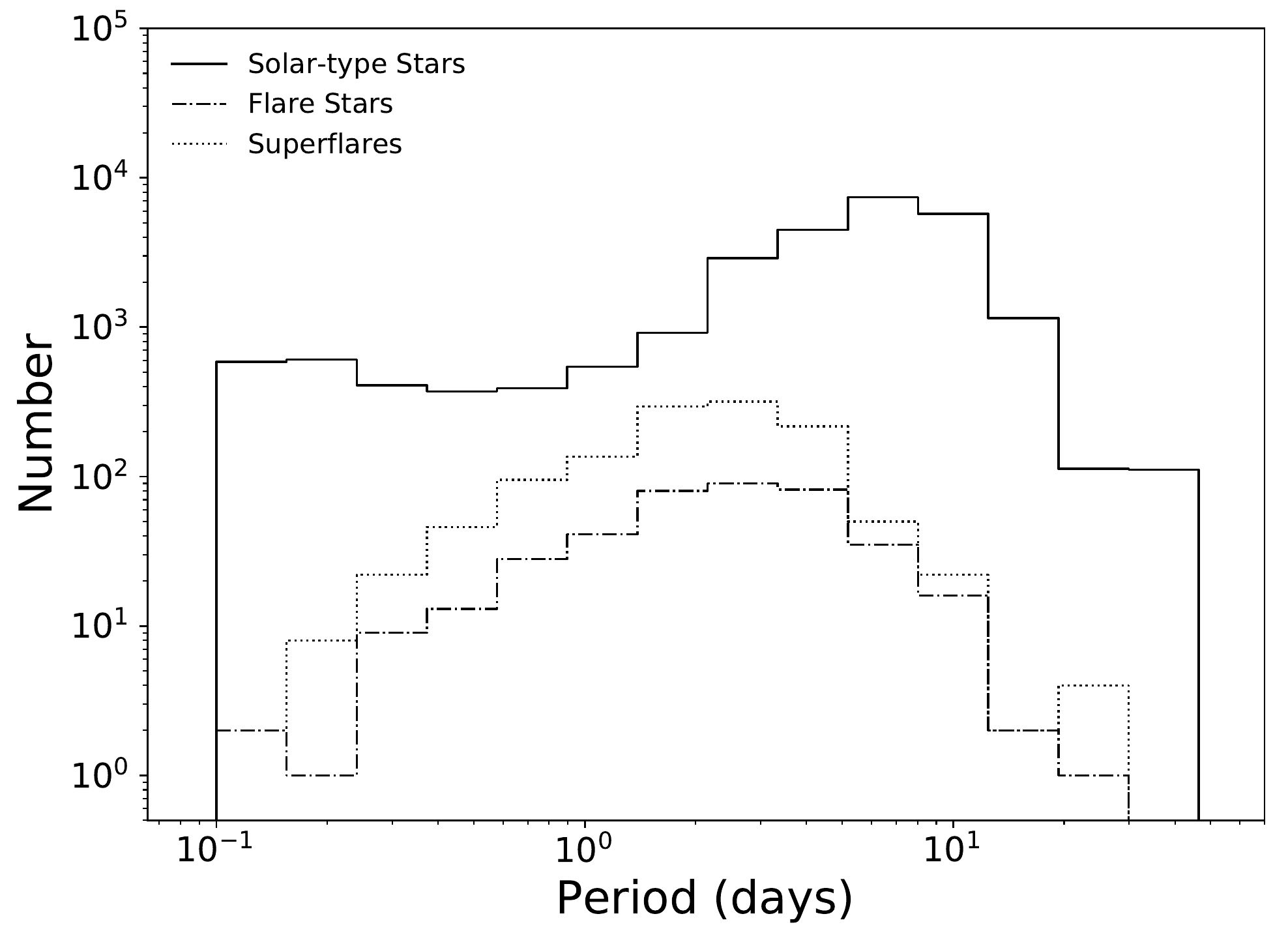}
\caption{Periodic distribution of solar-type stars (solid line),
superflares (dotted line), and flare stars (dashed line).
Table \ref{tab:num_period} lists the corresponding values of each
subset.}
        \label{fig:num_period}
    \end{figure*}

\begin{figure*}[!hp]
\centering
\includegraphics[width=0.6\linewidth]{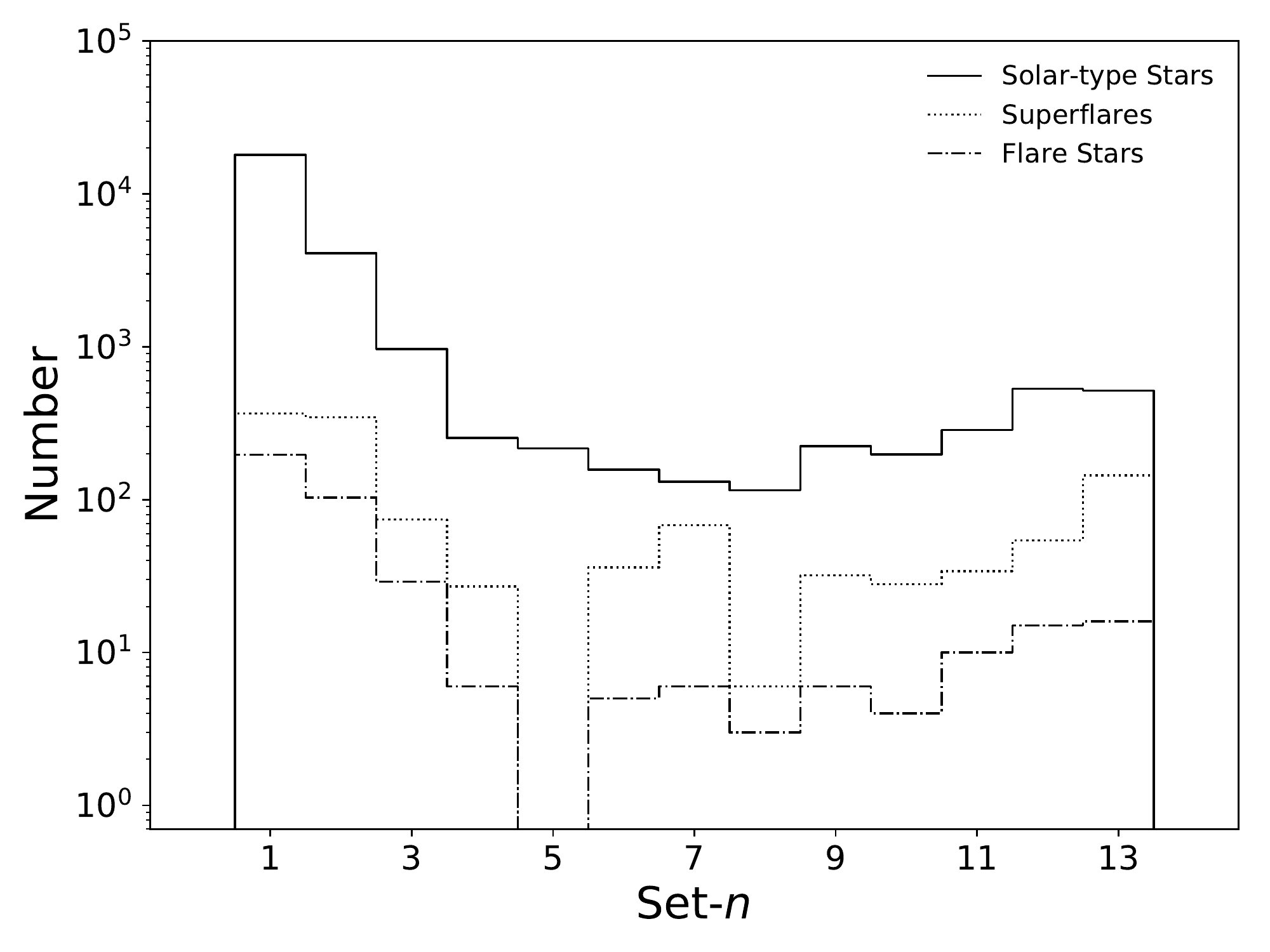}
\caption{Distribution of solar-type stars (solid line), superflares
(dotted line), and flare stars (dashed line) in each Set-$n$ subset.
The definition of Set-$n$ can be found in Section
\ref{sec:oc-fre-distri}. Table \ref{tab:SetN} lists the corresponding
values of each subset.}
\label{fig:SetN}
\end{figure*}

Figure \ref{fig:num_period}  present
the period distributions of the solar-type stars, superflares and
flare stars. Note that we may miss some slowly rotating solar-type
stars with $P>10$ days because of the limited observing span. Most
of the  {\em TESS} targets were observed only in one sector, i.e.
about 27 days, as shown in Figure \ref{fig:SetN}. Therefore, periods
longer than $\sim$14 days are not reliable for these targets.
In total, we obtain stellar periods for 3827 slowly rotating ($P>10$
days) solar-type stars and 22,207 fast rotating ($P<10$ days)
solar-type stars.

\subsection{Energy of superflares}
\label{sec:superflares-energy} Following the method of
\citet{2015ApJ...798...92W}, the stellar luminosity can be estimated
with the Stefan--Boltzmann law
\begin{equation}\label{equ:lumi}
L_{*}=4 \pi R_{*}^{2} \sigma_{\mathrm{sb}} T_{*}^{4},
\end{equation}
where $R_{*}$ and $T_{*}$ are the stellar radius and effective
temperature given by TIC v8, and $\sigma_{\mathrm{sb}}$ is the
Stefan--Boltzmann constant. The flare energy thus can be calculated
by integrating the fluxes within the flare event
\begin{equation}\label{equ:flare_energy}
E_{\text {flare }}=\int L_{*} F_{\text {flare }}(t) d t,
\end{equation}
where $F_{\text {flare }}(t)$ is the normalized flux
above the fitted quadratic function (see section
\ref{sec:superflares-selection})
\begin{equation}\label{equ:flare_flux}
F_{\text {flare }}(t) = F(t)-F_{\rm q}{(t)}.
\end{equation}

\section{Results}\label{sec:results}
In Table \ref{tab:Flare stars}, we list all parameters of 400 flare stars. Table \ref{tab:Flares} gives the parameters of 1216 superflares. Table \ref{tab:num_period} present
the period distributions of the solar-type stars, superflares and
flare stars.
\subsection{Occurrence frequency distribution}
\label{sec:oc-fre-distri}

The observation mode of  {\em TESS}, unlike {\em Kepler}, causes
various observing spans for different targets. It is not suitable for
calculating the occurrence frequency of superflares directly using the
unequal observing spans. We therefore improve the method suggested
by \citet{2012Natur.485..478M}. First of all, we subdivided all the
solar-type stars into different sets based on how many sectors the
star was observed in. For example, Set-${1}$ means that the stars
were observed in only one sector. Similarly, Set-${13}$ covers the
stars observed in all the 13 sectors. The count of the solar-type
stars in each Set-$n$ can be found in Figure \ref{fig:SetN} and
Table \ref{tab:SetN}.

For each Set-$n$, the superflare frequency distribution as a function of flare energy is defined as

\begin{equation}\label{equ:frequency}
f_{n}=\frac{N_{\text {flares}}}{N_{\text {stars}} \cdot \tau_{ {n}}
\cdot \Delta E_{\text {flare}}},
\end{equation}
where $N_{\text {flares}}$ and $N_{\text {stars}}$ are the numbers
of superflares and
solar-type stars in Set-$n$. $\Delta E_{\text{flare}}$ represents the bin width of flare energy. $\tau_n$ can be
calculated as
\begin{equation}\label{equ:tau_n}
\tau_{ {n}} = n \times 23.4 \; \text{days}.
\end{equation}
As the observing time is affected by satellite orbit, {\em TESS
} may not observe fully and effectively observe for 27 days in each sector.
Here, we calculated the mean value of continuous observation length
of each sector, which is $23.4$ days. The final occurrence
frequency of superflares for all the solar-type stars can be
calculated as
\begin{equation}\label{equ:13frequency}
f=\frac{{\sum}f_{n}}{13}.
\end{equation}

Figure \ref{fig:oc_fre}(a) illustrates the distribution of
flare peak amplitudes, i.e. the peak flux of $F_{\rm flare}(t)$ in
Equation (\ref{equ:flare_flux}). Panel (b) presents the frequency
distributions as a function of flare energy for all the solar-type
stars (solid line), and slowly (dashed line) or rapidly (dotted
line) rotating solar-type stars. After comparing the dotted and
dashed lines, it is evident that rapidly rotating stars are much more
active than slowly rotating stars. According to the relation between
stellar periodicity and age
\citep{1972ApJ...171..565S,2003ApJ...586..464B}, this result
therefore indicates that younger solar-type stars generate
superflares more frequently.

A power-law model is used to fit the frequency distributions

\begin{equation}
\frac{d N }{d E}\propto E^{-\gamma}.
\end{equation}
The power-law index $\gamma$ is fitted by using the same linear
regression method as in \citet{2018ApJ...869L..23T}. For all solar-type
stars, $\gamma$ is $ 2.16 \pm 0.10$, which is consistent with
$\gamma \sim 2.2$ from \citet{2013ApJS..209....5S} within a 1$\sigma$
interval, but higher than $\gamma \sim 1.5$ from
\citet{2015EP&S...67...59M}. From Figure
\ref{fig:oc_fre}(c), it is obvious that hotter stars (with $5600
\mathrm{K} \leqslant T_{\mathrm{eff}}<6000 \mathrm{K}$) have lower
frequency than cooler stars (with $5100 \mathrm{K} \leqslant
T_{\mathrm{eff}}<5600 \mathrm{K}$). The above results are basically
similar to those found from {\em Kepler} data
\citep{2012Natur.485..478M,2013ApJS..209....5S,2015EP&S...67...59M}.

Figure \ref{fig:oc_fre}(d) compares the flare frequency derived in this work and by
\citet{2015EP&S...67...59M}, which used the 1 minute short-cadence
data of  {\em Kepler} and therefore detected more flares in the
low-energy region ($< 2\times10^{34}$ erg). In the high-energy
region ($> 2\times10^{34}$ erg), our superflare frequencies are
higher than theirs. We speculate that this result is caused by the
higher proportion of young stars in our dataset, as we have
discussed above about (panel (b)). Since young stars rotate more
rapidly and are more active, they are more likely to generate
superflares. In total, 25,734 solar-type stars are selected, among
which 21,955 stars have periods of less than 10 days. The proportion of
young stars is 85\% in our work, compared to 32\% in the work by
\citet{2015EP&S...67...59M}. One may refer to Appendix
\ref{sec:app_period} for more discussions about the potential
changes caused by the number fraction of rapidly and slowly rotating
stars.

The superflare frequency of Sun-like stars (dotted line) is also
presented in Figure \ref{fig:oc_fre}(d). For Sun-like
stars (with $5100 \mathrm{K} \leqslant T_{\mathrm{eff}}<6000
\mathrm{K}$ and $P>10$ days), we obtain a higher frequency of
superflares ($>10^{35} \rm erg$) than that from {\em Kepler}
\citep{2012Natur.485..478M,2013ApJS..209....5S,2015EP&S...67...59M}.
However, there are only seven such events, which are insufficient to
robustly make a statistical conclusion. We look forward to more
observations from {\em TESS} to extend the sample of Sun-like stars.

\subsection{Stellar period versus superflare energy }\label{sec:period_energy}
An apparent negative correlation between flare frequency and
stellar period has been found
\citep{2012Natur.485..478M,2013ApJ...771..127N,2015EP&S...67...59M}.
In Figure \ref{fig:period energy}(b), in the range of 
the stellar period over a few days, it is clear that the flare frequency
decreases with an increase of the rotation period. A similar result is
found by \citet{2013ApJ...771..127N} and
\citet{2019ApJ...876...58N}. Moreover, stellar age has positive
correlation with the stellar period
\citep{1972ApJ...171..565S,2003ApJ...586..464B}. This result
indicates that rapidly rotating stars, or young stars, are more
likely to generate superflares.

\citet{2019ApJ...876...58N} found that the upper limit of flare
energy in each period bin has a continuous decreasing trend with the
rotation period. From Figure \ref{fig:period energy}(a), compared
with the previous result \citep[Figure 12(b)
of][]{2019ApJ...876...58N}, this trend is not obvious for the whole
range of periods. However, superflares within the period range over a
few days (the tail part) clearly show the decreasing trend. We
separate superflares into two parts according to their surface
effective temperatures, which are shown in Figure \ref{fig:period
energy} (c) and (d) respectively. Superflares with a temperature $5600
\mathrm{K} \leqslant T_{\mathrm{eff}}<6000 \mathrm{K}$ perhaps slightly more clear decreasing trend for the whole period range,
compared with panel (a). A similar decreasing trend can also be found
in these two panels with period range over a few days (the tail
part). As period is the key factor, we look forward to other methods
precisely determining the periods of these solar-type stars
(Appendix \ref{sec:app_period}). Meanwhile, slow-rotation stars
($P\sim 25$ days) need to be searched for specially, in order to
statistically and strongly confirm this trend.

\subsection{Active flare stars}\label{sec:activ_star}
Given that solar-type stars are grouped in different Set-$n$ because
of unequal observing spans, the flare frequency for an individual
star is described by
\begin{equation}\label{equ:activefrequency}
f_{*}=\frac{N_{\text {*flares}}}{\tau_{ {*}}},
\end{equation}
where $\tau_{ {*}}$ is the continuous observation length of each flare star, and $N_{\rm
{*flares}}$ denotes the number of flares from an individual star.

Table \ref{tab:starfrequency} and Table \ref{tab:Nstarflares} list
the top 20 stars sorted by flare frequency $f_{*}$ and the number of
flares $N_{\text {*flares}}$, respectively. TIC43472154 is the most
active star with an impressive flare frequency. If it is not
coincidentally observed at an extraordinary active period,
TIC43472154 can generate 233.17 flares per year. Besides, this
star is not accompanied by any fainter stars or M dwarfs, according
to the cross-match results from {\em Gaia} and {\em Hipparcos}
data. This occurrence rate is much higher than that of KIC10422252,
which is 41.6 flares per year \citep{2013ApJS..209....5S}.  Figure
\ref{fig:TIC_43472154} plots the light curve of TIC43472154.
Superflares are marked with downward arrows.

Another target, TIC364588501, was observed by all sectors of {\em
TESS} and exhibits the largest number of flares (63).
Additionally, like TIC43472154, this star is also a single
star, according to the cross-match results from {\em Gaia} and {\em
Hipparcos} data. TIC364588501 became more active than usual in the
last 15 days of sector 5, and generated 15 superflares (Figure
\ref{fig:TIC_364588501}).

The most energetic superflare comes from TIC93277807, releasing
$1.77\times 10^{37} {\rm erg}$ in around 1.5 hr. This value consists
with the inference that energy releasing through stellar flare is
saturated at $\sim 2\times 10^{37} {\rm erg}$
\citep{2015ApJ...798...92W}.

\subsection{Planets of flare stars}\label{sec:planet}
Hot Jupiters are considered as one of the essential factors that can
produce superflares on host stars
\citep[e.g.][]{2000ApJ...529.1031R,2004ApJ...602L..53I}. According
to these studies, it is not possible for the Sun to generate
superflares. However, a lot of statistical works based on {\em
Kepler} data did not detect hot Jupiters around flare stars
\citep{2012Natur.485..478M,2013ApJS..209....5S,2015EP&S...67...59M}.

We made a cross-check between our 400 flare stars with the Exoplanet
Follow-up Observing Program for {\em TESS} (ExoFOP--{\em
TESS}\footnote{https://exofop.ipac.caltech.edu/tess/}). Cross-
matching results are listed in Table \ref{tab:planets}.  There are
four stars (TIC25078924, 44797824, 257605131, and 373844472) that
have planet candidates. These four stars are unlikely to affect our
statistical results. We thus did not exclude them in the analysis.
Besides, there is only one star (TIC410214986) that possesses a
confirmed hot planet \citep{2019ApJ...880L..17N}. This is an
interesting planet, and we list its parameters in Table
\ref{tab:planets}, even if its hosting star (TIC410214986) is
flagged as a binary system from the {\em Hipparcos}-2 catalog.

Compared with other stars, there is nothing special about the
superflares of these four targets. Therefore, our results suggest
that the planet-star interactions are unlikely to be a general
mechanism for producing superflares.

Impacts of stellar activities toward their hosted planets have
been studied comprehensively \citep[e.g.][]{2010AsBio..10..751S,
2016NatGe...9..452A, 2017MNRAS.465L..34A, 2017ApJ...848...41L}. Some
reviews have briefly introduced how superflares would affect their
hosted planets
\citep[e.g.][]{2018SSRv..214...21R,2019arXiv190505093A,2019LNP...955.....L}.
Planets around flare stars may not only help us to understand
planet-star interactions in generating flares, but also importantly
extend our knowledge about how superflares will affect the space weather
of related planets. This will definitely improve our understanding
about the relation between the Sun and the Earth as well. Many more
space- and ground-based missions will focus on searching habitable
planets. However their habitability is just simply decided by
habitable zones. From a more foresighted aspect, it is also
important and necessary to estimate habitability in detail by examining the impacts of their hosting stars' activities.

\subsection{Correlation between superflare energy and duration}\label{sec:ETcorrelation}
\citet{2015EP&S...67...59M} proposed that the energy and duration of
superflares are connected through power-law function, which was also
proved by \citet{2018ApJ...869L..23T} using the statistical testing
method. By applying magnetic reconnection theory, which has been
widely accepted as the mechanism of solar flares, the correlation
between the energy ($E$) and duration ($T_{\text {duration }}$) of
superflares can be denoted as
\begin{equation}\label{equ:energyduration}
T_{\text {duration }} \propto E^{1/3}.
\end{equation}

Benefiting from the 1 minute short-cadence data from {\em Kepler}, their
calculations of the duration and energy of superflares are estimated more accurately than when
using long-cadence data. The value of $\beta$ is found to be
$0.39\pm0.03$ \citep{2015EP&S...67...59M}.

The 2 minute cadence of {\em TESS} also makes it possible to acquire
accurate flare duration and energy. We use the superflares
detected in this paper and obtain $\beta=0.42 \pm 0.01$, which is a
little larger than the result from {\em Kepler}. Figure
\ref{fig:sflare} presents the strong correlation between the
duration and energy of superflares in our dataset.

\citet{2015EP&S...67...59M} concluded that the same physical mechanism
may be shared by both solar flares and superflares of solar-type
stars, according to their similarity. Later,
\citet{2017ApJ...851...91N} used solar white-light flares, and got
$\beta=0.38 \pm 0.06$, which is remarkably consistent with the
result $\beta=0.39\pm0.03$ of \citet{2015EP&S...67...59M}. This
similarity may support the idea that stellar superflares and solar
flares are both generated by magnetic reconnections. But stellar
superflares and solar white-light flares cannot be fitted by just one
single power-law function, which indicates their potential
difference. By using data from {\em Galaxy Evolution
Explorer} (GALEX) missions, \citet{2019ApJ...883...88B} found that
short-duration and near-ultraviolet flares do not show a strong
correlation between energy and duration. Here, the observational
values of $\beta$ derived from {\em Kepler} and {\em TESS} are both
larger than the theoretical prediction $\beta =1/3$. Since this
value is totally derived by the theory, which is successfully
applied to solar flares, it may not precisely illustrate superflares
on solar-type stars.

We consider the magnetic field strength of the ${\rm Alfv\acute{e}n}$
velocity ($v_{A}=B / \sqrt{4 \pi \rho}$) as a variable, and assume the preflare coronal density as a constant. The scaling
power-law relation of equation (\ref{equ:energyduration}) can be
expressed as
\begin{equation}\label{equ:T-E-B}
T_{\text {duration }} \propto E^{1/3}B^{-5/3},
\end{equation}
\begin{equation}\label{equ:T-E-L}
T_{\text {duration }} \propto E^{-1/2}L^{5/2},
\end{equation}
where $B$ is the magnetic field strength and $L$ is the flaring length
scale. For a more detailed introduction of these two relations, one may
refer to \citet{2017ApJ...851...91N}. Using the coefficients
obtained by \citet{2017ApJ...851...91N}, we also plot these two
scaling laws in Figure \ref{fig:sflare}(b). Meanwhile, solar
white-light flares \citep{2017ApJ...851...91N} and superflares of
solar-type stars found by using {\em Kepler} short-cadence data
\citep{2015EP&S...67...59M} are also presented in this figure.

From Figure \ref{fig:sflare}, the majority of superflares found in this
work (blue circles) have a magnetic field strength around 60--200G
and length scales near $10^{10} $--$10^{11} {\rm cm}$. 
The magnetic strength of superflares in this work are lower than that of
superflares found by using {\em Kepler} short-cadence data (red
squares). As we discuss in Section \ref{sec:DM} and Appendix
\ref{sec:KvsT}, our method of selecting superflares based on
pixel-level data and the bandpass of {\em TESS} tends to exclude weak
flares with faint flare signal, which means that the superflares in
this work have a higher energy compared to those of {\em
Kepler} (from Figure \ref{fig:Mflare}(c)). Apparently, superflares
have a higher magnetic strength and a larger flaring length than solar white-light flares do. Superflares with lower energies tend to be
generated from magnetic fields with a higher strength and  a smaller
flaring length scale. In contrast, those superflares with higher
energies tend to be from the weaker magnetic fields and larger flaring length scales. In the future, the
observation of stellar magnetic strength and the improvement of
photometric imaging for precisely measuring flaring length scales
will definitely advance our understanding of superflares. Also, by
including more parameters of the superflares under consideration, a
close-to-reality physical pattern will describe superflares in
detail.

\section{Summary}
\label{sec:conclusions} In this work, 25,734 solar-type stars are
selected from the first year's observations by {\em TESS} in the southern
hemisphere of the sky. For the first time, we detect 1216 superflares from 400
flare stars by using 2 minutes cadence data from {\em TESS}.

Statistical research of these superflares is applied. We
calculate the occurrence frequency distribution as a function of
superflare energy. The power-law index $\gamma$ of the superflare
frequency distribution ($dN/dE\propto E^{-\gamma}$) is constrained
to be $\gamma = 2.16\pm 0.10$, which is consistent with the results
from {\em Kepler}.

According to statistics from all solar-type stars, hotter stars
(with $5600 \;\mathrm{K} \leqslant T_{\mathrm{eff}}<6000 \; \mathrm{K}$)
have lower superflare frequency than cooler stars (with $5100\;
\mathrm{K} \leqslant T_{\mathrm{eff}}<5600 \;\mathrm{K}$). Besides,
rapidly rotating stars ($P<10$ days) are more active than
slowly rotating stars ($P>10$ days). These two conclusions are
basically consistent with those of the {\em Kepler} data.

It is worth mentioning that the frequency distributions calculated
by using {\em TESS} data are, overall, higher than those from the {\em
Kepler} data. The reason is that the majority of the solar-type
stars in our dataset are rapidly rotating stars with stellar
periodicities shorter than 10 days. Moreover, as only seven superflares
are detected on Sun-like stars, we may not be able to give a
convincing conclusion on the frequency of superflares for Sun-like
stars.

Some solar-type stars are very interesting in our dataset. For
example, TIC43472154 is the most active star and has an occurrence
rate of 233.17 superflares per year. TIC364588501 exhibits 63
superflares during the first year of the {\em TESS} mission.
Additionally, several stars may have planet candidates. TIC410214986
has been confirmed to possess a hot planet, although this star is
flagged as a binary system by {\em Hipparcos}. Finally, the
correlation between the energy and duration of superflares is
calculated. The power-law indexes from {\em TESS} ($\beta =
0.42\pm0.01$) and {\em Kepler} ($\beta = 0.39\pm0.03$) are both
larger than the theoretical value ($\beta=1/3$), which is derived
from the magnetic reconnection theory of solar flares.

A dataset containing superflares of solar-type stars has been
constructed using {\em TESS}'s first year observations for 
the first time, given that {\em TESS} stars are bright and near, which
is convenient for ground observations. In the future, photometric
and spectrometric observations should be introduced for further
studies of superflares. On the one hand, it is important to exclude
those binary stars. Meanwhile, evolved stars (e.g. red giants and
subgiants) potentially included in this work should be removed with
newly spectrometric measurements of their surface temperatures and
radii. Other methods (see Appendix \ref{sec:app_period}) should be
applied for the periodic estimation of flare stars, because the period is a
key parameter. High-resolution photometric observation will
definitely be helpful to obtain clean light curves without any
contamination from other, fainter, stars. On the other hand, for
superflares, spectrometric information will be useful in tracking
the chromospheric activities \citep{2016NatCo...711058K}. With
high-resolution photometric observations, it will be easier to
determine whether superflares come from the primary star. The Zwicky
Transient Facility (ZTF) will simultaneously observe the northern
hemisphere with {\em TESS} \citep{2019RNAAS...3..136V}. Compared with
the 21$''$ pixel scale of {\em TESS}, the resolution of ZTF is 1$''$ per pixel. The telescope will make nightly g-- and r--band
observations, which will offset the deficiency of the {\em TESS} bandpass on
400--600nm. It may observe some less energetic flares with
high-resolution imaging. Now, {\em TESS} is targeting the northern
celestial hemisphere, which makes it possible to use the whole-sky
data for studying superflares, and it is enlarging the dataset of
superflares on Sun-like stars.

\section*{Acknowledgements}
We thank an anonymous referee for constructive suggestions. We
would like to thank Z.G. Dai, Y.F. Huang, X.F. Wu, H. Yu, B.B. Zhang,
and G.Q. Zhang for helpful discussions. N. Liu gave us helpful
suggestions about cross matching works with data from {\em Gaia}-DR2
and the {\em Hipparcos}-2 catalog. We sincerely thank the Mikulski Archive
for Space Telescopes (MAST) and the TESS community for applying
convenient data portal and tools. This work is supported by the
National Natural Science Foundation of China (grants
U1831207,11803012).

\appendix
\section{{\em TESS} versus {\em Kepler}}\label{sec:KvsT}
Here, we note our considerations according to the instrument
features of {\em TESS}. The pixel scale ($\sim 21''$) of {\em
TESS} is five times larger than that ($\sim 4''$) of {\em
Kepler}. Around the whole sky, {\em TESS} attempts to search for
planets transiting bright and nearby stars, which make it convenient
for the ground-based follow-up observation. Therefore, the pixel scale of {\em
TESS} may not need to be as high resolution as it is for{\em Kepler}. But
the larger pixel scale may also contain other energetic activities
which are not from the primary objects, or the star may actually be a
binary star.

We first exclude binary stars in our dataset by using the {\em
Hipparcos}-2 catalog (HIP2) \citep{2007A&A...474..653V}. One thing
that should be noticed is that when we cross match HIP2, only 3113 stars in
our dataset contain the {\em Hipparcos} identifier and the remaining
solar-type stars still cannot be fully treated as single solar-type
stars. They should be further confirmed by follow-up observations. As
145 binary stars only consist of 4.7\% of 3113 stars, even if we exclude
all binary stars, the flare frequencies may not be significantly
changed.

Then, we use {\em Gaia}-DR2 \citep{2018A&A...616A...1G} to
search for other unrelated stars which are located in the 42$''$ radius of
the primary solar-type stars. We exclude 155 stars, of which
brighter stars located within 21$''$. In other words, at just
one pixel scale their flux will indistinguishably contaminate the
primary objects. We have searched for some M dwarfs within a 42$''$ radius, these are distributed around with 2849 solar-type stars.
We do not exclude them from the data set for the following reasons:
(1) M dwarfs should have $3000 \mathrm{K} < T_{\mathrm{eff}}<4000
\mathrm{K}$, and $\log g>4.0$ \citep[e.g.][]{2017ApJ...849...36Y}.
{\em Gaia}-DR2 only provides surface effective temperature, but the
surface gravity $\log g$ is not included in the catalog. Therefore,
the searched targets are just candidates of M dwarfs. (2) Flare
frequencies of M dwarfs are much larger than those of solar-type
stars (a.k.a. G dwarfs). But their flare energy ranges from $\sim
10^{31}$ to $10^{36}$ erg \citep{2017ApJ...849...36Y}, and 95\% of
flares have energies of $10^{32.40\pm 1.35}$ erg. This energy range
is much smaller than superflares on solar-type stars. Therefore, even if M
dwarfs are in the same aperture, the possibility that their flares
contaminate the data set of this work is much lower. We demonstrate this
idea in Figure \ref{fig:Mflare}(a). In the figure, histograms of
flare energy in a $\log$ scale of flares on M dwarfs
\citep{2019MNRAS.489..437D} and superflares on solar-type stars
(this work) are shown. Solid curves give the fitting results by
normal distribution. Flares on M dwarfs from
\citet{2019MNRAS.489..437D} are detected also by using {\em TESS}
data. Apparently, according to the energy distributions of these two
kinds of flares, superflares on solar-type stars are two orders of
magnitude higher than flares on M dwarfs. 

The bandpass filters for {\em TESS} and {\em Kepler} are
different, 600--1000nm for {\em TESS} \citep{2015JATIS...1a4003R} and 420--900nm for {\em Kepler} \citep{2016Kepler}. Comparing
these two bandpass filters, {\em TESS} is more sensitive to longer
wavelengths, which are designed to detect many more M dwarfs.
White-light flares detected from other stars can be described by
blackbody radiation with effective temperature in the range of 9000 K -- 10,000 K
\citep[e.g.][]{2003ApJ...597..535H,2010ApJ...714L..98K}, of which
the peak emission is considered to be blue. Therefore,
\citet{2019MNRAS.489..437D} have argued that according to the
bandpass of {\em TESS}, it did not detect less energetic flares on M
dwarfs. This idea is proved in Figure \ref{fig:Mflare}(b). In the
figure, flare energy distributions of M dwarfs are from
\citet{2017ApJ...849...36Y} (using {\em Kepler} data) and
\citet{2019MNRAS.489..437D}(using {\em TESS} data), which are
colored by red and green, respectively. It is obvious that the
flares detected by {\em TESS} have relatively higher energies than
those from {\em Kepler}. 

Similarly, in Figure \ref{fig:Mflare}(c), we compare energy
distributions of superflares on solar-type stars that are detected
by {\em Kepler} (colored in magenta) and {\em TESS} (colored in
blue). Note that superflares detected by {\em Kepler}
\citep{2013ApJS..209....5S} may not totally be generated on
main-sequence G-type dwarfs. So, first of all, we cross match the
dataset of \citet{2013ApJS..209....5S} with the another catalog, in
which radii of {\em Kepler} stars are revised by using {\em
Gaia}-DR2 \citep{2018ApJ...866...99B}. Then, 496 superflares from
main-sequence dwarfs are left. From the figure, we find that the
superflares from our dataset have relatively higher energies than
those from {\em Kepler} data. Therefore, {\em TESS} is more likely
to detect relatively energetic flares. The same indication can also be
found in Figure \ref{fig:oc_fre}, where there is a low frequency of superflares with energies lower than $10^{35}$erg. These results are mainly
caused by the incomplete detection of flares with low energies. 

\section{The Number Fractions of Solar-type Stars in an Equation of Stellar Periods}\label{sec:app_period}
One thing that should be kept in mind is that there are potential
differences between the observed and real number fractions of slowly
and rapidly rotating stars. Following Appendix B of
\cite{2019ApJ...876...58N}, we use the empirical gyrochronology
relation to simply estimate the number fraction of solar-type stars
in an equation of period. This empirical relation can be written as
\citep[e.g.][]{2008ApJ...687.1264M}
    \begin{equation}\label{equ:NpNall}
    N_{\mathrm{star}}\left(P_{\mathrm{rot}} \geqslant P_{0}\right)=\left[1-\frac{t_{\mathrm{gyro}}\left(P_{0}\right)}{\tau_{\mathrm{MS}}}\right] N_{\mathrm{all}},
    \end{equation}
where $t_{\mathrm{gyro}}\left(P_{0}\right)$ represents the stellar
gyrochronological age in the equation of period, which can be estimated
by Equation (12)-(14) of \citet[][]{2008ApJ...687.1264M}. The
corresponding $B-V$ values of different effective temperatures are
estimated by Equation (3) of \citet{2012EL.....9734008B}.
$\tau_{\mathrm{MS}}$ is the main-sequence phase, and here we set it
as a constant ($\sim 10~{\rm Gyr}$) for all observation fields of
{\em TESS}. One can refer to Appendix B of
\cite{2019ApJ...876...58N} for more specific calculation details.
Here, we just list our results in Table \ref{tab:N10result}. 

\begin{figure*}[!hp]
    \gridline{\fig{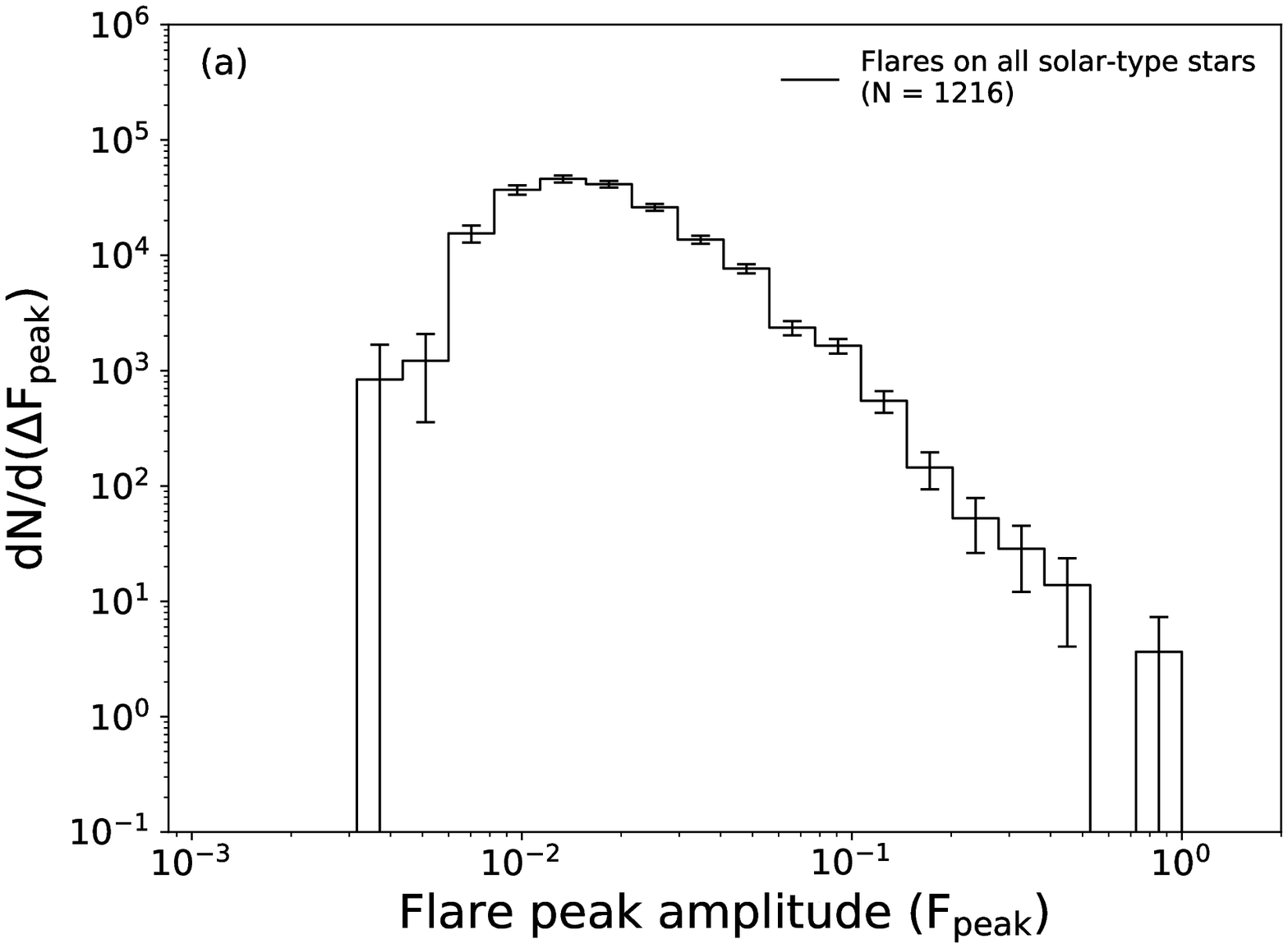}{0.5\textwidth}{}
        \fig{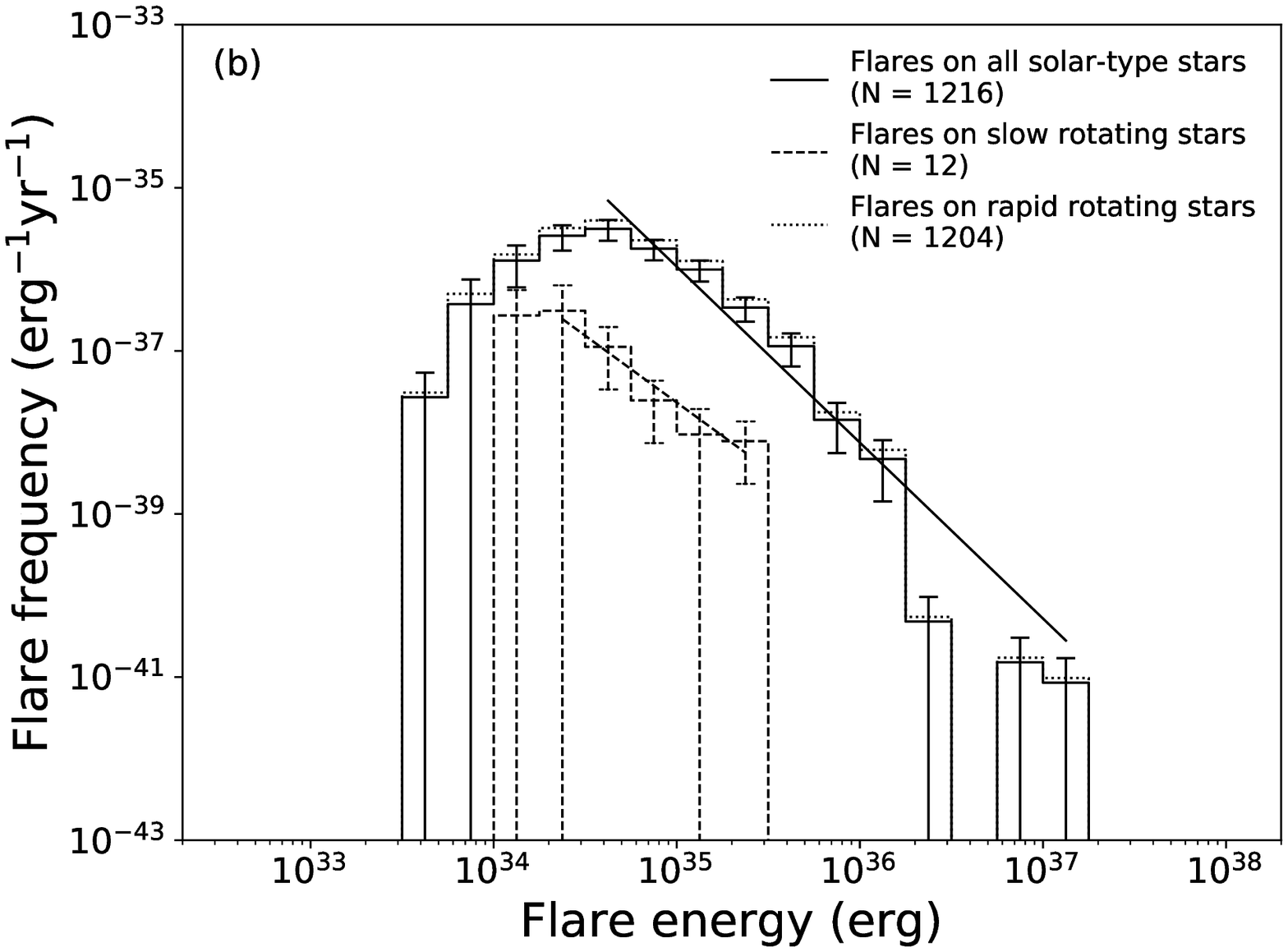}{0.5\textwidth}{}
    }
    \gridline{\fig{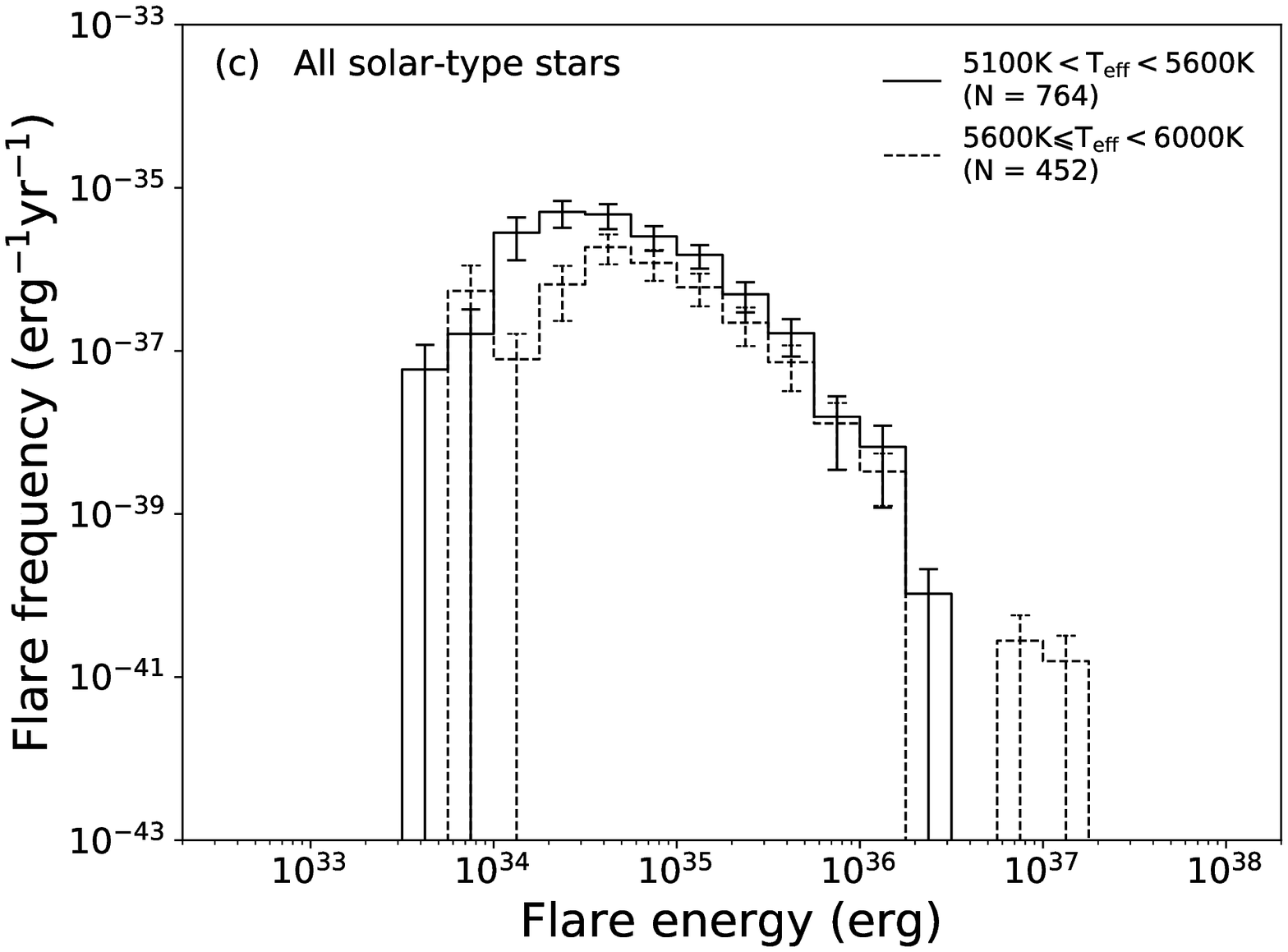}{0.5\textwidth}{}
        \fig{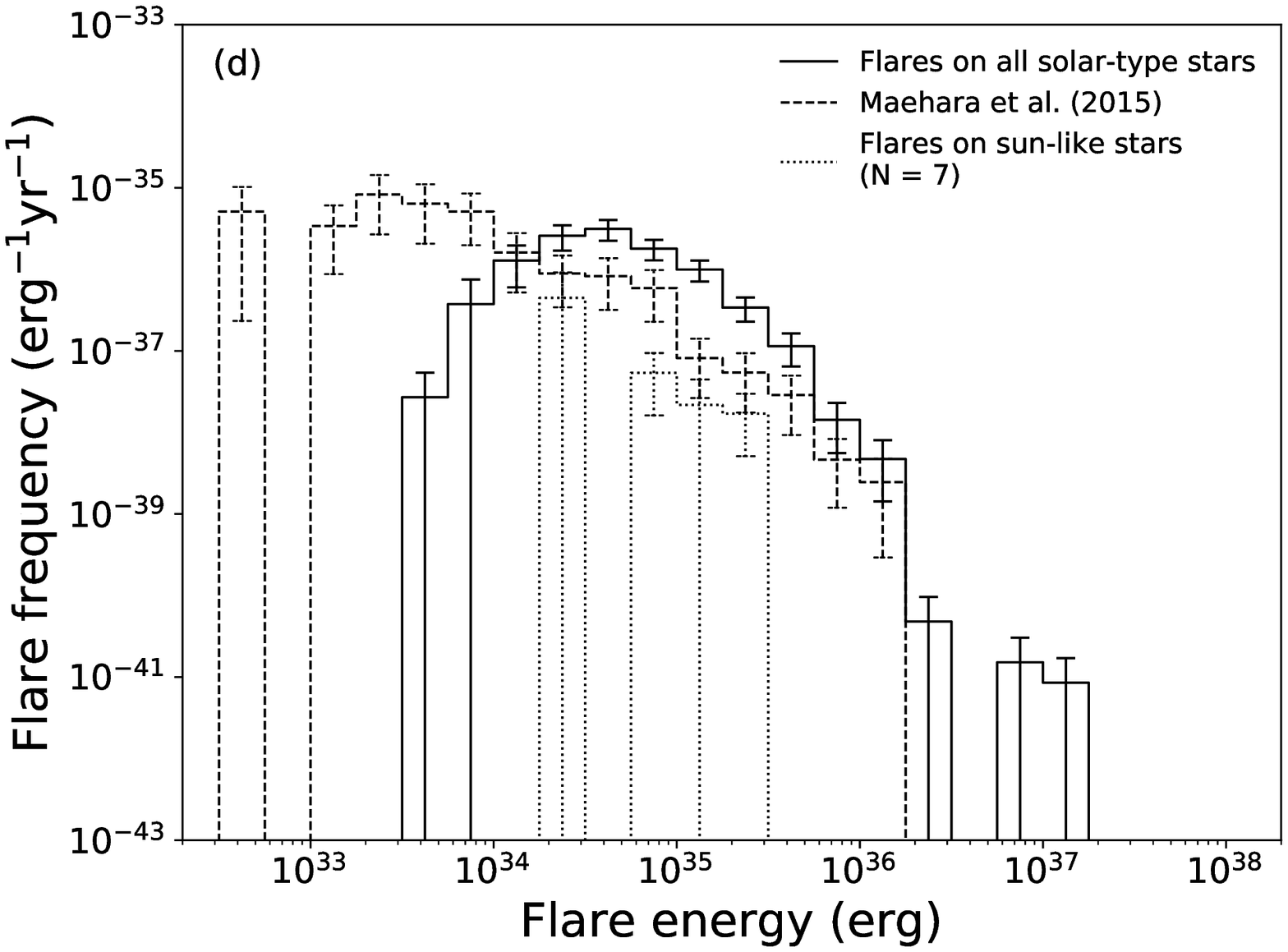}{0.5\textwidth}{}
    }
    \caption{
        (a) Frequency distribution of the normalized peak
flux $F_{\rm peak} =F(t_{\rm peak}) - F_q(t_{\rm peak})$ at time $t_{\rm peak}$. All 1216 superflares are included.
(b) Frequency distribution of flare energy. The solid line represents all
1216 solar-type stars (with $5100 \mathrm{K} \leqslant
T_{\mathrm{eff}}<6000 \mathrm{K}$ and $\log g>4.0$), and the dashed line
expresses 12 superflares of slowly rotating solar-type stars (with a period $P>10$ days). Two straight lines show the best-fit power-law
distribution of $d N / d E \propto E^{-\gamma}$. The solid line gives
$\gamma \sim 2.16 \pm 0.10$, and the dashed line gives $\gamma \sim 1.64
\pm 0.44$. The dotted line stands for 1204 flares on rapidly rotating
stars, of which frequency distributions are almost overlap with the solid line.
(c) The same frequency distribution as in (b) but for different
effective surface temperatures. The solid and dashed lines represent
data sets of $5100 \mathrm{K} \leqslant T_{\mathrm{eff}}<5600
\mathrm{K}$ and $5600 \mathrm{K} \leqslant T_{\mathrm{eff}}<6000
\mathrm{K}$ respectively.
(d) The solid line indicates the same frequency
distribution as the solid line in (b).
In contrast with the result of {\em
Kepler} short-cadence data, we imported the result from Figure \ref{fig:num_period}(b) of
\citet{2015EP&S...67...59M} as the dashed line. The dotted line
indicates the frequency distribution of Sun-like stars with an effective
temperature $5600 \mathrm{K} \leqslant T_{\mathrm{eff}}<6000
\mathrm{K}$ and period $P>10$ days. However, only seven flares satisfy the two criteria.
In all panels, the error bars are
given by the square root of the flare numbers in each bin.
            \label{fig:oc_fre}}
\end{figure*}

\begin{figure*}[!h]
    \centering
    \gridline{\fig{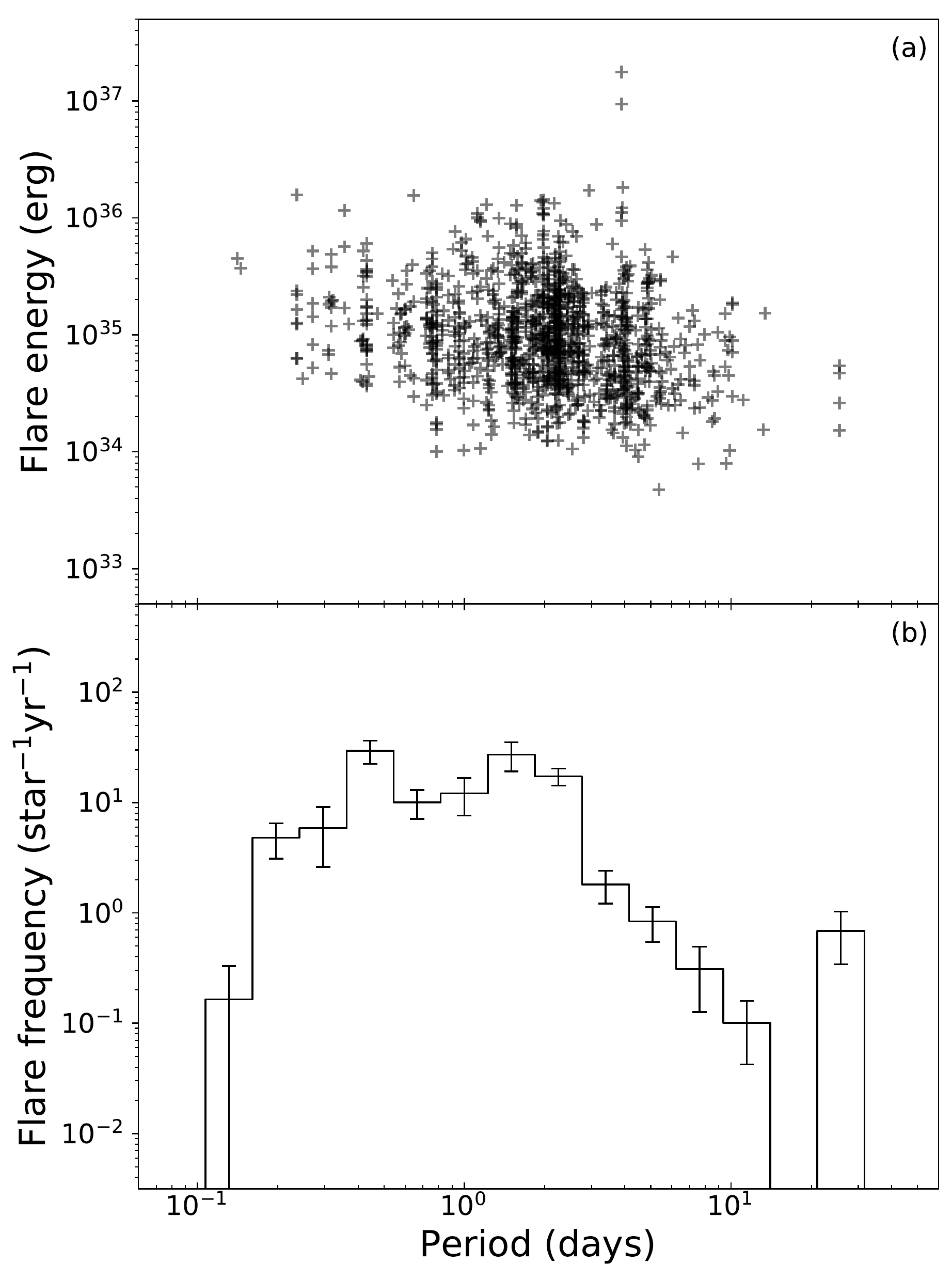}{0.49\textwidth}{}
        \fig{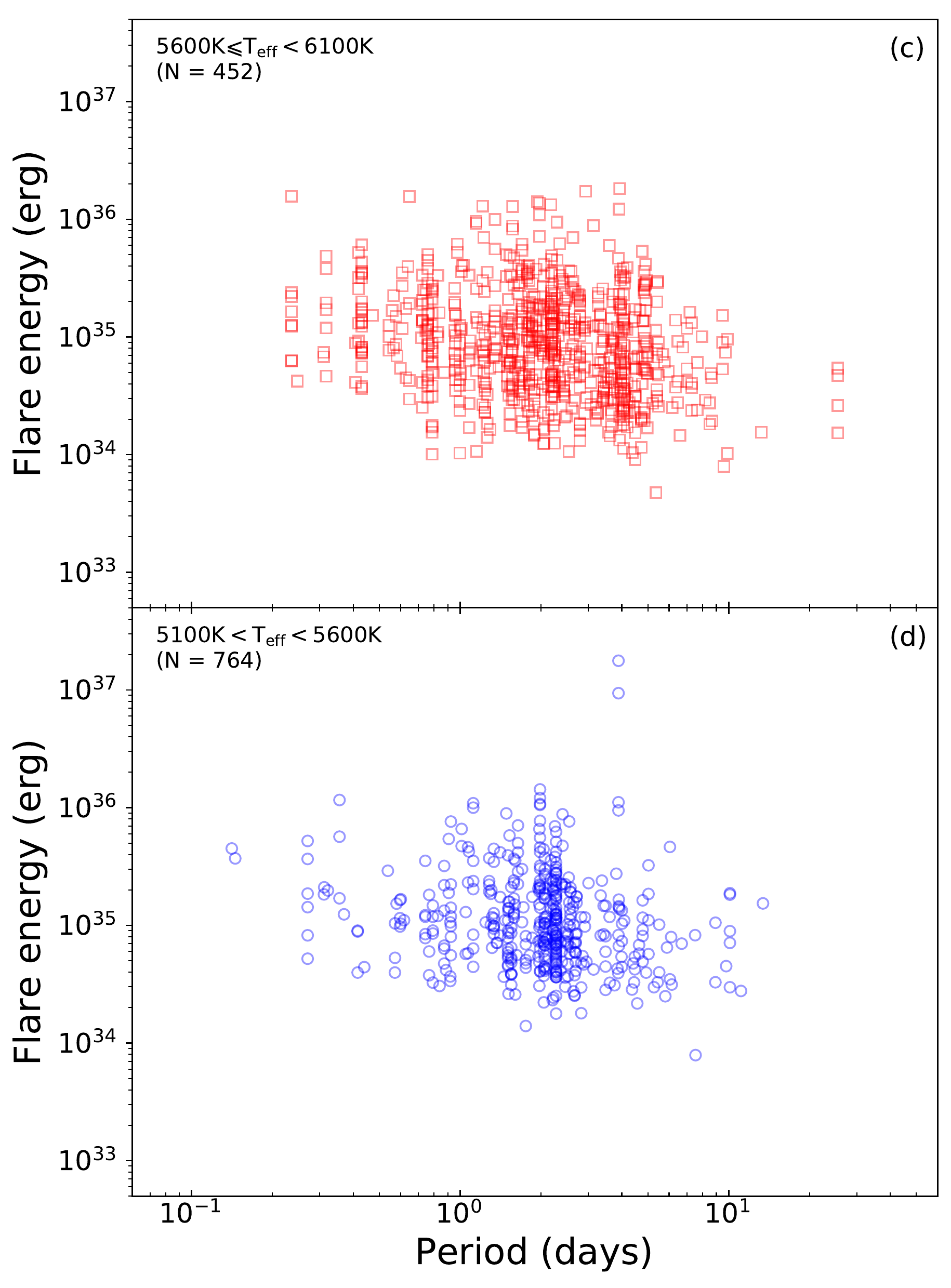}{0.49\textwidth}{}
    }
    \caption{(a) Scatter plot of superflare energy vs. stellar period.
        Black pluses denote every single superflare.
        (b) Flare frequency of each period bin; the error bars are
        given by the square root of flare numbers in each bin.
        (c), (d) Same as panel (a), but stand for superflares
        with different effective temperatures
        $5600 \mathrm{K} \leqslant T_{\mathrm{eff}}<6000 \mathrm{K}$ (red squares) and
        $5100 \mathrm{K} \leqslant T_{\mathrm{eff}}<5600 \mathrm{K}$ (blue circles), respectively.}
    \label{fig:period energy}
\end{figure*}

\begin{figure*}[!tp]
        \centering
        \includegraphics[width=1\linewidth]{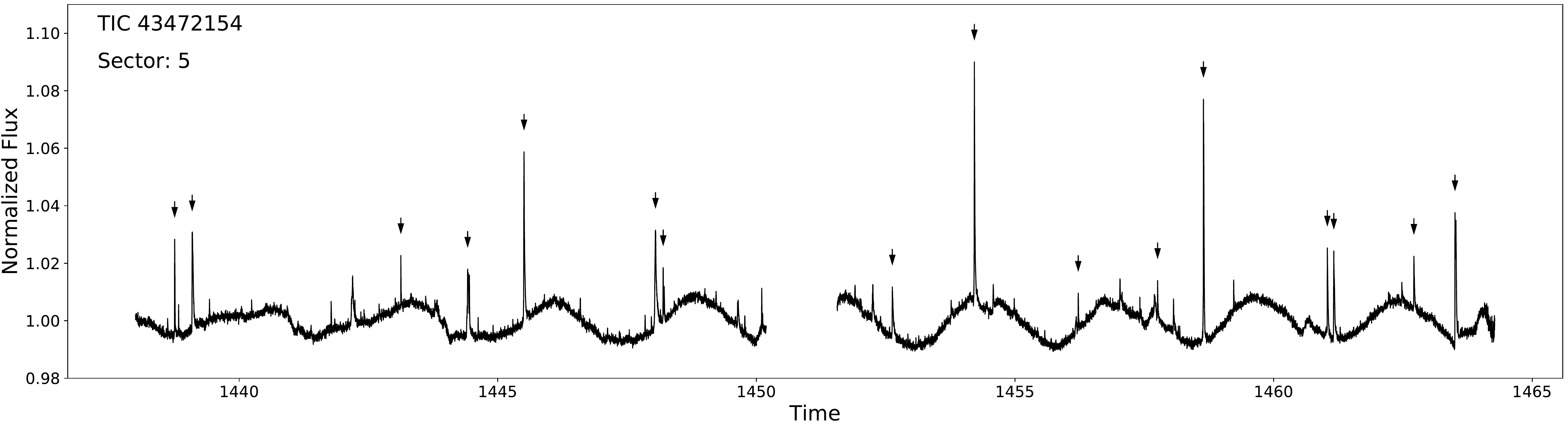}
\caption{Light curve of TIC43472154. Small arrows mark those
superflares selected from automatic software, and checked through a pixel-level pattern.}
        \label{fig:TIC_43472154}
    \end{figure*}

\begin{figure*}
        \centering
        \includegraphics[width=1\linewidth]{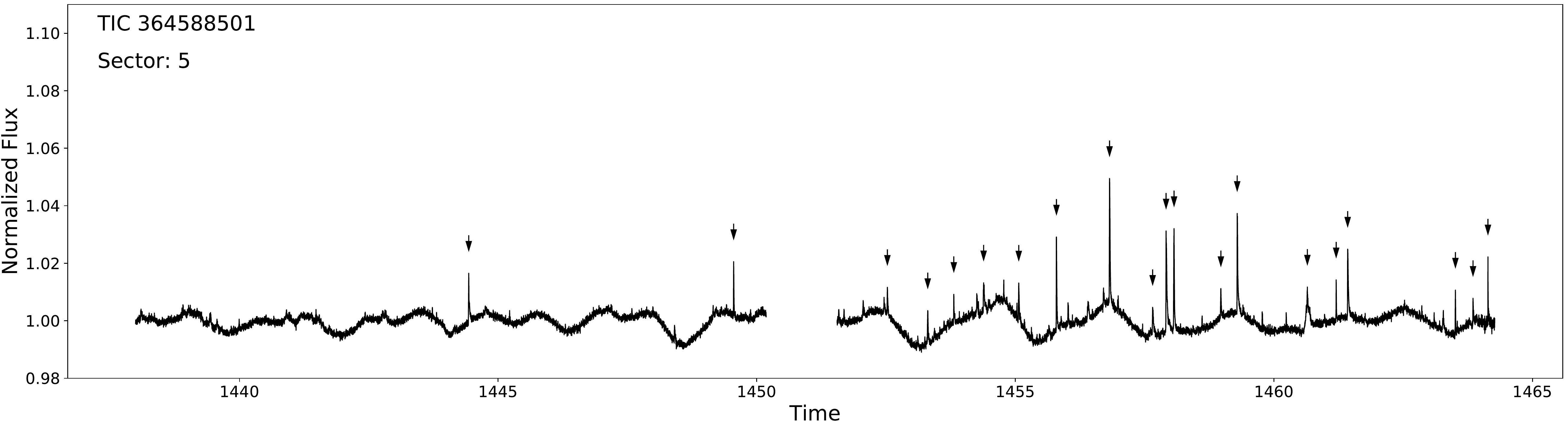}
        \caption{Same as Figure \ref{fig:TIC_43472154}, but for TIC364588501.}
        \label{fig:TIC_364588501}
    \end{figure*}

\begin{figure*}
        \centering
        \gridline{\fig{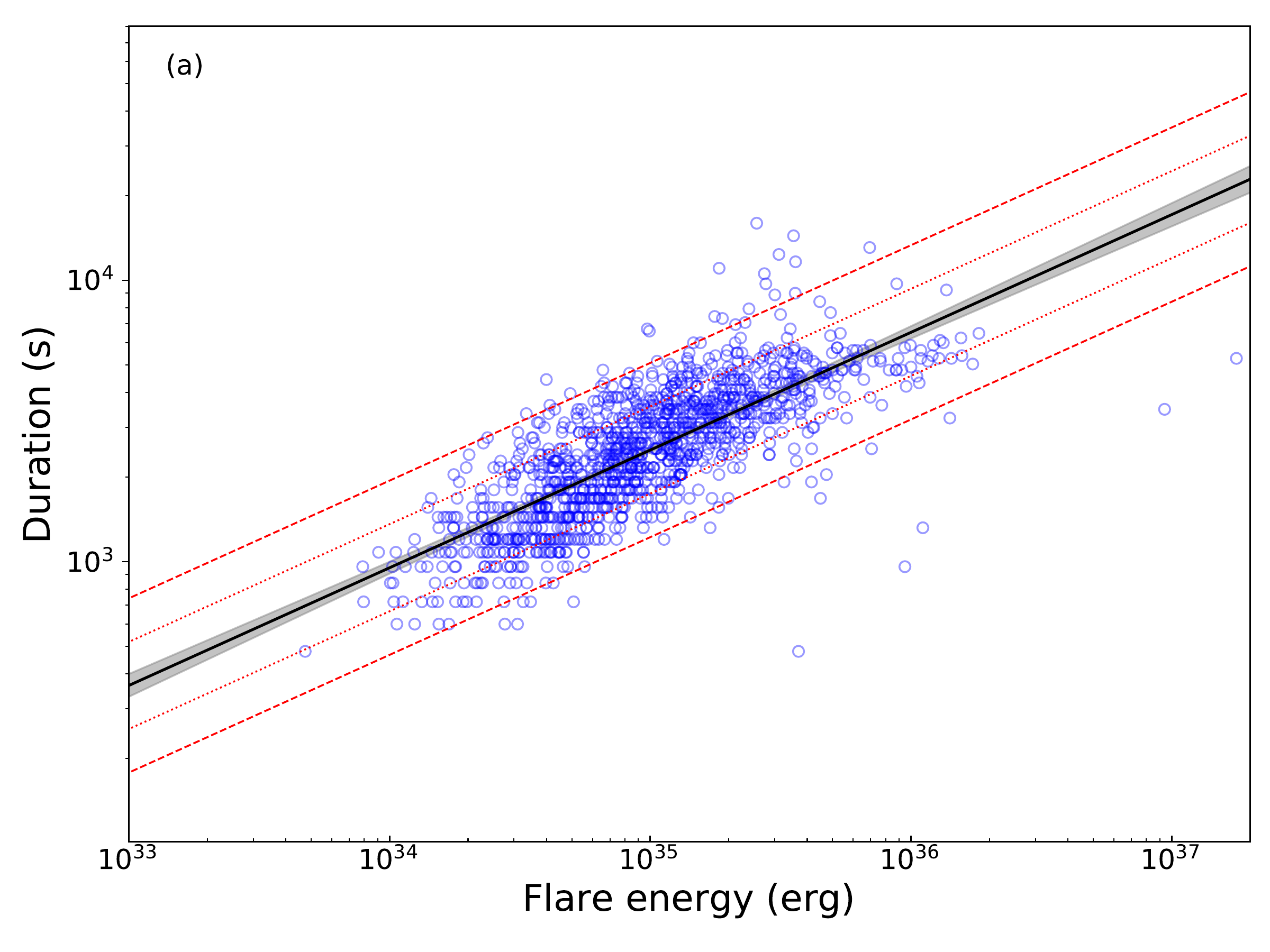}{0.49\textwidth}{}
            \fig{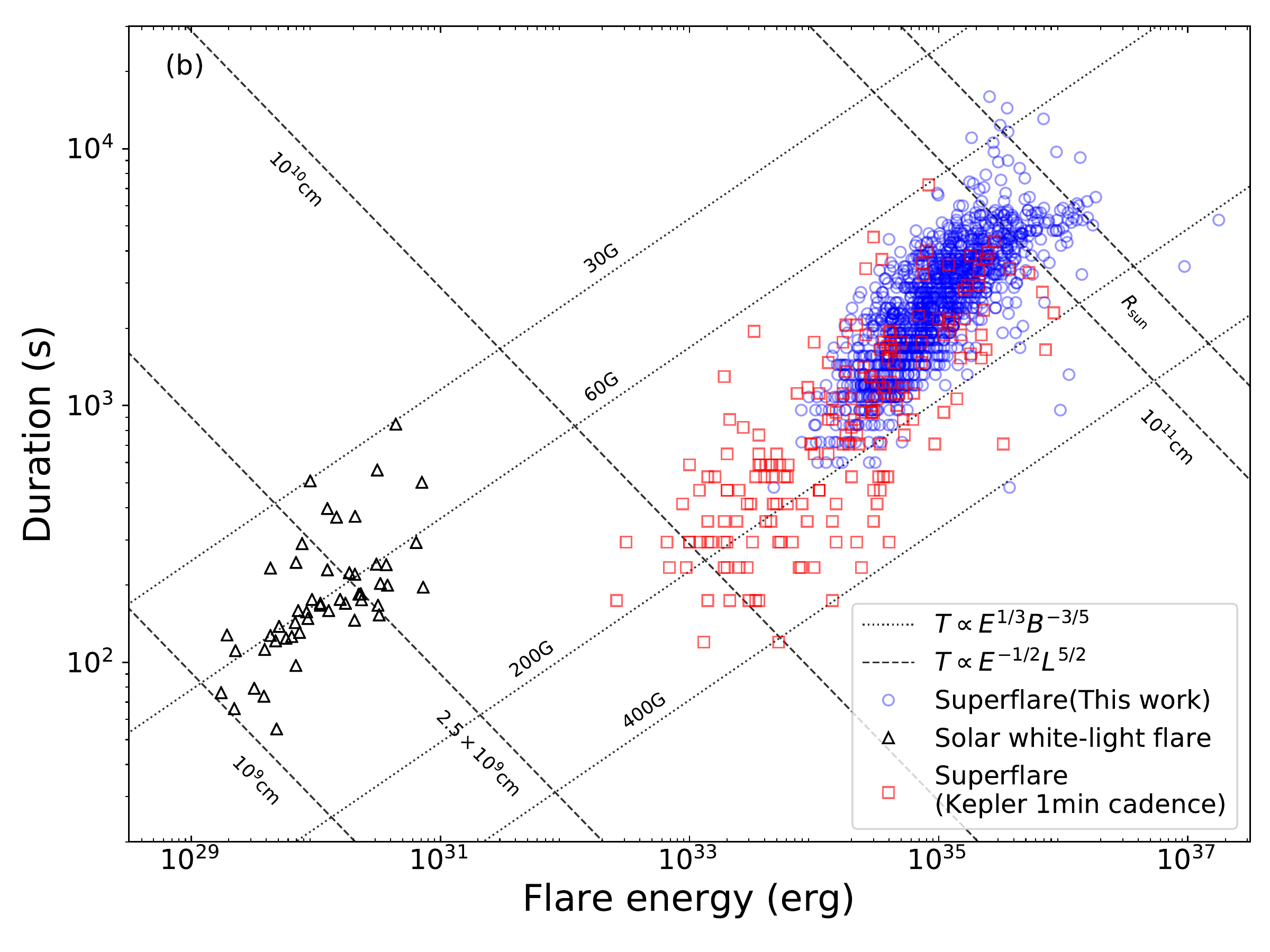}{0.49\textwidth}{}
        }
\caption{(a) Correlation between superflare energy and duration.
    The best fitting is marked with black solid line.
    The gray area represents the 95\% confidence interval of fitting
    uncertainties. The red dotted and dashed lines
    represent $1\sigma$ and $2\sigma$ intervals of extra variability,
    which are denoted as $\sigma_{v}$ in Section 3.1 of
    \citet{2018ApJ...869L..23T}.
    (b) Comparing superflares in this work (blue circles) with solar white-light flares (black triangles; \citet{2017ApJ...851...91N}) and superflares found by using the short-cadence
    data from {\em Kepler} (red squares; \citet{2015EP&S...67...59M}).
    The dotted and dashed lines represent scaling laws of Equation (\ref{equ:T-E-B}) and
    Equation (\ref{equ:T-E-L}), respectively. The coefficients for plotting these lines
    are the same as in \citet{2017ApJ...851...91N}. Here, $B$ represents
    magnetic field strength, and $L$ denotes flaring length scale.
}   \label{fig:sflare}
    \end{figure*}

\begin{figure*}
    \centering
    \includegraphics[width=0.6\linewidth]{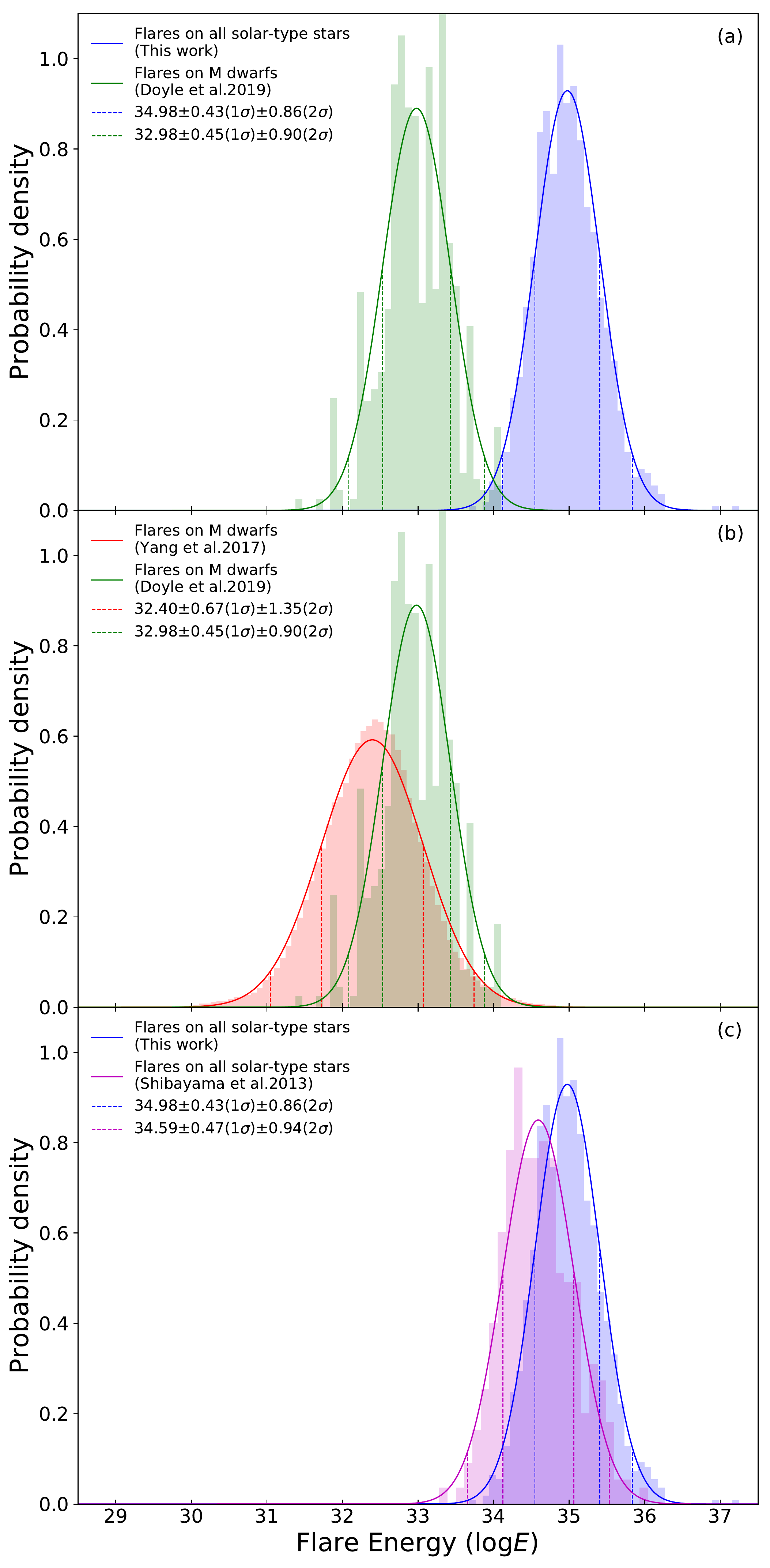}
    \caption{Energy distributions of superflares on solar-type stars
    (\citet[][]{2013ApJS..209....5S} and this work) and flares on M dwarfs \citep{2017ApJ...849...36Y,2019MNRAS.489..437D}. (a) The blue histogram shows the results from the superflares on solar-type stars (this work), and green shows flares on M dwarfs \citep{2019MNRAS.489..437D}. These two works both use {\em TESS} data. (b) Flares on M dwarfs are marked in red, and from {\em Kepler} data \citep{2017ApJ...849...36Y}, and the green histogram is the same as in (a). (c) The magenta histogram represents the
    superflares on solar-type stars from \citet{2013ApJS..209....5S}, which used data from {\em Kepler}. The blue histogram is the as same in (a). The corresponding results of fitting the normal distribution are all listed in the legend of each panel, and shown by the solid colored lines. Dashed lines give standard deviations in ranges of $1\sigma$ and $2\sigma$.
}\label{fig:Mflare}
\end{figure*}

From the table, we may notice that the number fraction of
solar-type stars with periods over 10 days period in our dataset is five to
seven times less than the roughly estimated values, and five to six
times less than the results of the {\em Kepler} field. Even though the
flare frequency of slowly rotating stars in Figure
\ref{fig:oc_fre}(b) and the flare frequency in the equation of period in
Figure \ref{fig:period energy}(b) can be five to seven times
smaller, the dependency to the period may not be changed significantly,
as these changes are less than one order of magnitude. While considering more solar-type stars with periods
over 10 days, the number of flares should also statistically
increase. Therefore, the changes are much smaller than they look. From
another aspect, the results roughly estimated by empirical relations may
not be reliable \citep[e.g.][]{2015A&A...577L...3T,
2016Natur.529..181V}. We set the main-sequence phase as a constant
for all {\em TESS} fields, which is not scientifically strict. 

In the future, the main-sequence phase can
be studied precisely for the field of each camera in each sector of
{\em TESS} observation. Although the estimation of longer stellar
periods is hard under the limitations of {\em TESS}, we hope other
methods can be applied to precisely determine the periodicity of {\em
TESS} stars. For example, through the autocorrelation function (ACF)
\citep[e.g.][]{2014ApJS..211...24M}, long periods can be determined
by measuring the rotational velocity by stellar spectrum
\citep[e.g.][]{2008ApJ...684.1390R,2010AJ....139..504B,2012AJ....143...93R,2018A&A...614A..76J}.
 As {\em TESS} is targeting the northern hemisphere,
those stars which were observed by {\em Kepler} and {\em TESS} can
use period estimation from the light curves of {\em Kepler} or another
{\em Kepler} periodic catalog \citep[e.g.][]{2014ApJS..211...24M}.


\begin{table*}[!h]
    \renewcommand{\arraystretch}{1}
    \addtolength{\tabcolsep}{+1pt}
    \caption{Flare stars}
    \label{tab:Flare stars}
    \centering
    \begin{tabular}{lccccrrcc}
        \hline \hline
        \multicolumn{1}{c}{{\em TESS} ID} & $T_{\rm eff}$ $^a$ & $\log{g}$ $^b$ & Radius $^c$   & Period $^d$ & \multicolumn{1}{c}{Flares $^e$} & \multicolumn{1}{c}{Set-$n$ $^f$} & $f_{\rm *}$ $^g$    & Flag $^h$ \\
        \multicolumn{1}{c}{}                               & (${\rm K}$)              &                & ($R_{\odot}$) & (days)      & \multicolumn{1}{c}{}            & \multicolumn{1}{c}{}             & (${\rm year}^{-1}$) &           \\ \hline
        737327                                             & 5872               & 4.20           & 1.35          & 1.62        & 1                               & 1                                & 14.57               & GM        \\
        1258935                                            & 5739               & 4.48           & 0.97          & 4.93        & 1                               & 1                                & 19.37               &           \\
        6526912                                            & 5541               & 4.24           & 1.25          & 2.29        & 1                               & 1                                & 14.39               &           \\
        7491381                                            & 5692               & 4.47           & 0.96          & 2.21        & 4                               & 3                                & 21.37               &           \\
        7586485                                            & 5801               & 4.36           & 1.11          & 1.76        & 6                               & 2                                & 46.11               &           \\
        11046349                                           & 5409               & 4.31           & 1.12          & 3.33        & 2                               & 1                                & 29.48               & GM        \\
        12359032                                           & 5471               & 4.51           & 0.90          & 2.32        & 2                               & 1                                & 27.98               &           \\
        12393800                                           & 5571               & 4.19           & 1.31          & 1.03        & 1                               & 1                                & 13.99               &           \\
        13955147                                           & 5701               & 4.44           & 1.01          & 2.22        & 1                               & 2                                & 7.69                &           \\
        15444490                                           & 5598               & 4.09           & 1.49          & 1.98        & 2                               & 2                                & 15.85               &           \\ \hline
    \end{tabular}
    \begin{flushleft}
        \textsc{Note.}\\ {
            $^a$ Effective surface temperature in unit of ${\rm K}$. \\
            $^b$ Surface gravity of star in $\log$ scale.\\
            $^c$ Stellar radii in unit of solar radius $R_{\odot}$.\\
            $^d$ Stellar periods in units of days.\\
            $^e$ Number of flares of each star (same as $N_{\rm*flares}$ of Equation (\ref{equ:activefrequency})).\\
            $^f$ Number of Set-$n$ which defined in Section \ref{sec:oc-fre-distri}.\\
            $^g$ Flare frequency deduced by Equation (\ref{equ:activefrequency}).\\
            $^f$ Flags of flare stars, where GM means the star may possess  M dwarf candidates nearby (42$''$ from the main target). GB indicates that there are stars that are brighter than the main stars, and 21--42$''$ from the main targets.\\
        (A portion of data is shown here. The full data is available in machine-readable form online.)}\\
        \vspace{1.5em}
    \end{flushleft}

\end{table*}

\begin{table*}[!h]
    \renewcommand{\arraystretch}{1}
    \addtolength{\tabcolsep}{+1pt}
    \caption{Superflares}
    \label{tab:Flares}
    \centering
    \begin{tabular}{ccccr}
        \hline \hline
        \multicolumn{1}{c}{{\em TESS} ID} & Peak Date $^a$ & Peak Flux $^b$ & Energy $^c$ & Duration $^d$           \\
        &                & (erg s$^{-1}$)        & (erg)       & \multicolumn{1}{c}{(s)} \\ \hline
        737327                         & 1460.1798      & 5.02E+31       & 5.58E+34    & 1919.98                 \\
        1258935                        & 1517.6755      & 4.71E+31       & 3.97E+34    & 1560.03                 \\
        6526912                        & 1673.3840       & 3.78E+32       & 9.46E+35    & 5759.94                 \\
        7491381                        & 1469.0019      & 1.10E+32        & 7.96E+34    & 1559.99                 \\
        7491381                        & 1470.5130      & 4.08E+31       & 9.09E+34    & 3359.98                 \\
        7491381                        & 1486.3100      & 3.71E+31       & 2.32E+34    & 959.98                  \\
        7491381                        & 1489.2516      & 6.92E+31       & 2.12E+35    & 6959.85                 \\
        7586485                        & 1411.2694      & 5.53E+31       & 6.91E+34    & 2640.06                 \\
        7586485                        & 1428.4168      & 3.61E+31       & 4.75E+34    & 2520.01                 \\
        7586485                        & 1440.9098      & 2.07E+31       & 1.39E+34    & 959.99                  \\
        7586485                        & 1442.7237      & 9.31E+31       & 5.07E+34    & 1559.99                 \\
        7586485                        & 1447.3084      & 3.15E+31       & 4.37E+34    & 2279.97                 \\
        7586485                        & 1457.0429      & 3.82E+31       & 8.09E+34    & 3599.92            \\
        \hline
    \end{tabular}
    \begin{flushleft}
        \textsc{Note.}\\ {
            $^a$ The corresponding date of the superflares' peak. \\
            $^b$ Flux of peak calculated by $L_{*} F_{\rm {flare }}(t)$, when $t$ equals to the peak time. $L_{*}$ and $F_{\rm {flare }}(t)$ are defined in Equation \ref{equ:lumi} and Equation \ref{equ:flare_flux}.\\
            $^c$ Energy of the superflare.\\
            $^d$ Duration of the superflare in seconds.\\
        (A portion of data is shown here. The full data is available in machine-readable form online.)}\\
    \end{flushleft}

\end{table*}

\begin{table*}[!h]
        \renewcommand{\arraystretch}{1}
        \addtolength{\tabcolsep}{+10pt}
        \caption{Numbers of Solar-type stars, superflares, and Flare stars of each periodic Bin.}
        \label{tab:num_period}
        \centering
        \begin{tabular}{cccc}
            \hline \hline
            $\log \rm{P}$ & Solar-type Stars & Superflares & Flare Stars \\ \hline
            -1.00 & 585 & 2 & 2 \\
            -0.81 & 606 & 8 & 1 \\
            -0.62 & 408 & 22 & 9 \\
            -0.43 & 372 & 46 & 13 \\
            -0.24 & 390 & 95 & 28 \\
            -0.05 & 543 & 136 & 41 \\
            0.14 & 919 & 295 & 80 \\
            0.33 & 2900 & 317 & 90 \\
            0.52 & 4486 & 217 & 82 \\
            0.71 & 7410 & 50 & 35 \\
            0.90 & 5728 & 22 & 16 \\
            1.10 & 1151 & 2 & 2 \\
            1.29 & 113 & 4 & 1 \\
            1.48 & 111 & 0 & 0 \\
            1.67 & 7 & 0 & 0 \\
            1.86 & 3 & 0 & 0 \\
            2.05 & 2 & 0 & 0\\ \hline
        \end{tabular}
        \begin{flushleft}
            \textsc{Note.}\\ {$\log P$ represents stellar periodicity in the $\log$ scale.}\\
            \vspace{1.5em}
        \end{flushleft}

    \end{table*}

\begin{sidewaystable*}[!h]
        \caption{Numbers of Solar-type stars, Superflares, and Flare stars of Each Set-$n$.}
        \label{tab:SetN}
        \centering
        \begin{tabular}{crrrrrrrrrrrrrrrrrrr}
            \hline \hline
            \multirow{3}{*}{Set-$n$} & \multicolumn{7}{c}{$5100 \mathrm{K} \leqslant T_{\mathrm{eff}}<5600 \mathrm{K}$} & \multicolumn{1}{c}{} & \multicolumn{7}{c}{$5600 \mathrm{K} \leqslant T_{\mathrm{eff}}<6000 \mathrm{K}$} & \multicolumn{1}{c}{} & \multicolumn{3}{c}{\multirow{2}{*}{Total}} \\ \cline{2-8} \cline{10-16}
            & \multicolumn{3}{c}{$P<10$ days} & \multicolumn{1}{c}{} & \multicolumn{3}{c}{$P>10$ days} & \multicolumn{1}{c}{} & \multicolumn{3}{c}{$P<10$ days} & \multicolumn{1}{c}{} & \multicolumn{3}{c}{$P>10$ days} & \multicolumn{1}{c}{} & \multicolumn{3}{c}{} \\ \cline{2-4} \cline{6-8} \cline{10-12} \cline{14-16} \cline{18-20}
            & \multicolumn{1}{c}{$N_{\rm star}$} & \multicolumn{1}{c}{$N_{\rm flare}$} & \multicolumn{1}{c}{$N_{\rm fstar}$} & \multicolumn{1}{c}{} & \multicolumn{1}{c}{$N_{\rm star}$} & \multicolumn{1}{c}{$N_{\rm flare}$} & \multicolumn{1}{c}{$N_{\rm fstar}$} & \multicolumn{1}{c}{} & \multicolumn{1}{c}{$N_{\rm star}$} & \multicolumn{1}{c}{$N_{\rm flare}$} & \multicolumn{1}{c}{$N_{\rm fstar}$} & \multicolumn{1}{c}{} & \multicolumn{1}{c}{$N_{\rm star}$} & \multicolumn{1}{c}{$N_{\rm flare}$} & \multicolumn{1}{c}{$N_{\rm fstar}$} & \multicolumn{1}{c}{} & \multicolumn{1}{c}{$N_{\rm star}$} & \multicolumn{1}{c}{$N_{\rm flare}$} & \multicolumn{1}{c}{$N_{\rm fstar}$} \\ \hline
            1  & 7023 & 249 & 128 &  & 1204 & 0 & 0 &  & 8798  & 113 & 67  &  & 1018 & 5 & 2 &  & 18043 & 367  & 197 \\
            2  & 1465 & 233 & 67  &  & 406  & 0 & 0 &  & 1894  & 113 & 36  &  & 335  & 0 & 0 &  & 4100  & 346  & 103 \\
            3  & 326  & 34  & 14  &  & 88   & 0 & 0 &  & 471   & 39  & 14  &  & 80   & 1 & 1 &  & 965   & 74   & 29  \\
            4  & 92   & 16  & 3   &  & 24   & 0 & 0 &  & 123   & 11  & 3   &  & 14   & 0 & 0 &  & 253   & 27   & 6   \\
            5  & 73   & 0   & 0   &  & 19   & 0 & 0 &  & 109   & 0   & 0   &  & 15   & 0 & 0 &  & 216   & 0    & 0   \\
            6  & 54   & 18  & 2   &  & 14   & 0 & 0 &  & 68    & 18  & 3   &  & 21   & 0 & 0 &  & 157   & 36   & 5   \\
            7  & 39   & 65  & 5   &  & 16   & 0 & 0 &  & 63    & 3   & 1   &  & 13   & 0 & 0 &  & 131   & 68   & 6   \\
            8  & 37   & 1   & 1   &  & 15   & 0 & 0 &  & 48    & 5   & 2   &  & 15   & 0 & 0 &  & 115   & 6    & 3   \\
            9  & 80   & 31  & 5   &  & 29   & 0 & 0 &  & 90    & 1   & 1   &  & 25   & 0 & 0 &  & 224   & 32   & 6   \\
            10 & 62   & 25  & 2   &  & 35   & 0 & 0 &  & 73    & 3   & 2   &  & 28   & 0 & 0 &  & 198   & 28   & 4   \\
            11 & 91   & 20  & 6   &  & 44   & 4 & 1 &  & 115   & 10  & 3   &  & 35   & 0 & 0 &  & 285   & 34   & 10  \\
            12 & 182  & 29  & 5   &  & 77   & 1 & 1 &  & 213   & 23  & 8   &  & 59   & 1 & 1 &  & 531   & 54   & 15  \\
            13 & 173  & 38  & 5   &  & 76   & 0 & 0 &  & 193   & 106 & 11  &  & 74   & 0 & 0 &  & 516   & 144  & 16  \\ \hline
            Total & 9697 & 759 & 243 &  & 2047 & 5 & 2 &  & 12258 & 445 & 151 &  & 1732 & 7 & 4 &  & 25734 & 1216 & 400 \\ \hline
        \end{tabular}
        \begin{flushleft}
            \textsc{Note.}\\ {The definition of Set-$n$ is in Section \ref{sec:oc-fre-distri}.
            Basically, according to the stellar surface temperature and rotation period, we classify solar-type stars into four
            categories for each Set-$n$. $N_{\rm star}$, $N_{\rm flare}$, and $N_{\rm fstar}$ represent the numbers of solar-type stars, flares, and flare stars, respectively. The total numbers are also listed in the last three columns and the last row.
            }\\
            \vspace{1.5em}
        \end{flushleft}

    \end{sidewaystable*}

\begin{table*}[!h]
        \renewcommand{\arraystretch}{1}
        \addtolength{\tabcolsep}{+1pt}
        \caption{Stellar properties of flare stars sorted by star activity $f_{*}$.}
        \label{tab:starfrequency}
        \centering
        \begin{tabular}{lccccrrc}
            \hline
            \hline
            \multicolumn{1}{c}{{\em TESS} ID} & $T_{\rm eff}$ $^a$ & $\log{g}$ $^b$ & Radius $^c$ & Period $^d$ & \multicolumn{1}{c}{Flares $^e$} & \multicolumn{1}{c}{Set-$n$ $^f$} & $f_{\rm *}$ $^g$ \\
            \multicolumn{1}{c}{} & ($K$) &  & ($R_{\odot}$) & (days) & \multicolumn{1}{c}{} & \multicolumn{1}{c}{} & (${\rm year}^{-1}$) \\ \hline
            43472154  & 5316 & 4.49 & 0.90 & 2.80 & 16 & 1  & 233.17 \\
            92845906  & 5634 & 4.17 & 1.37 & 2.07 & 22 & 2  & 175.12 \\
            53417036  & 5555 & 4.44 & 0.98 & 1.53 & 8  & 1  & 140.26 \\
            20096356  & 5458 & 4.46 & 0.95 & 0.79 & 16 & 2  & 127.34 \\
            175491080 & 5321 & 4.52 & 0.88 & 3.98 & 14 & 2  & 122.71 \\
            206592394 & 5597 & 4.54 & 0.88 & 3.43 & 8  & 1  & 111.91 \\
            38402758  & 5472 & 4.42 & 1.00 & 1.08 & 6  & 1  & 97.62  \\
            127311608 & 5515 & 4.23 & 1.25 & 0.96 & 5  & 1  & 93.24  \\
            382575967 & 5567 & 4.42 & 1.01 & 2.19 & 40 & 7  & 91.36  \\
            284789252 & 5789 & 4.41 & 1.05 & 0.87 & 6  & 1  & 88.45  \\
            92347098  & 5234 & 4.48 & 0.90 & 3.93 & 5  & 1  & 81.34  \\
            152346470 & 5844 & 4.14 & 1.44 & 1.97 & 5  & 1  & 81.08  \\
            78055898  & 5495 & 4.30 & 1.14 & 4.07 & 10 & 2  & 79.59  \\
            257644579 & 5916 & 4.38 & 1.11 & 1.51 & 10 & 2  & 76.86  \\
            364588501 & 5605 & 4.26 & 1.22 & 2.28 & 63 & 13 & 75.35  \\
            272456799 & 5874 & 4.14 & 1.46 & 2.18 & 4  & 1  & 74.60  \\
            21540586  & 5417 & 4.27 & 1.18 & 1.35 & 9  & 2  & 71.63  \\
            93277807  & 5706 & 4.17 & 1.37 & 3.89 & 4  & 1  & 70.15  \\
            32874669  & 5147 & 4.53 & 0.84 & 4.73 & 4  & 1  & 70.15  \\
            302116397 & 5531 & 4.39 & 1.05 & 0.95 & 5  & 1  & 67.93 \\ \hline
       \end{tabular}
   \begin{flushleft}
   \textsc{Note.}\\ {Top 20 flare stars sorted by flare frequency ($f_{\rm *}$). The headers of this table are the same as in Table \ref{tab:Flare stars}.
   }\\
   \vspace{0.1em}
\end{flushleft}

    \end{table*}

\begin{table*}[!h]
        \renewcommand{\arraystretch}{1}
        \addtolength{\tabcolsep}{+1pt}
        \caption{Stellar properties of flare stars sorted by number of flares.}
        \label{tab:Nstarflares}
        \centering
        \begin{tabular}{lccccrrc}
            \hline
            \hline
            \multicolumn{1}{c}{{\em TESS} ID} & $T_{\rm eff}$ $^a$ & $\log{g}$ $^b$ & Radius $^c$ & Period $^d$ & \multicolumn{1}{c}{Flares $^e$} & \multicolumn{1}{c}{Set-$n$ $^f$} & $f_{\rm *}$ $^g$ \\
            \multicolumn{1}{c}{} & ($K$) &  & ($R_{\odot}$) & (days) & \multicolumn{1}{c}{} & \multicolumn{1}{c}{} & (${\rm year}^{-1}$) \\ \hline
            364588501 & 5605 & 4.26 & 1.22 & 2.28 & 63 & 13 & 75.35  \\
            382575967 & 5567 & 4.42 & 1.01 & 2.19 & 40 & 7  & 91.36  \\
            149539114 & 5367 & 4.36 & 1.05 & 0.43 & 22 & 10 & 36.05  \\
            92845906  & 5634 & 4.17 & 1.37 & 2.07 & 22 & 2  & 175.12 \\
            260162387 & 5855 & 4.20 & 1.35 & 1.98 & 18 & 13 & 21.63  \\
            167163906 & 5465 & 4.55 & 0.86 & 0.76 & 18 & 12 & 23.44  \\
            279572957 & 5570 & 4.34 & 1.11 & 1.81 & 16 & 9  & 27.86  \\
            43472154  & 5316 & 4.49 & 0.90 & 2.80 & 16 & 1  & 233.17 \\
            20096356  & 5458 & 4.46 & 0.95 & 0.79 & 16 & 2  & 127.34 \\
            167574282 & 5291 & 4.53 & 0.86 & 1.70 & 14 & 13 & 16.73  \\
            175491080 & 5321 & 4.52 & 0.88 & 3.98 & 14 & 2  & 122.71 \\
            339668420 & 5395 & 4.47 & 0.93 & 4.84 & 12 & 6  & 32.34  \\
            219389540 & 5744 & 4.44 & 1.02 & 1.55 & 11 & 6  & 29.32  \\
            257644579 & 5916 & 4.38 & 1.11 & 1.51 & 10 & 2  & 76.86  \\
            78055898  & 5495 & 4.30 & 1.14 & 4.07 & 10 & 2  & 79.59  \\
            38827910  & 5304 & 4.53 & 0.86 & 3.66 & 10 & 11 & 14.14  \\
            348898049 & 5226 & 4.71 & 0.69 & 1.00 & 9  & 13 & 10.76  \\
            219212899 & 5217 & 4.55 & 0.83 & 4.06 & 9  & 7  & 20.53  \\
            21540586  & 5417 & 4.27 & 1.18 & 1.35 & 9  & 2  & 71.63  \\
            260268898 & 5186 & 4.58 & 0.80 & 2.25 & 8  & 9  & 13.97 \\ \hline
        \end{tabular}
        \begin{flushleft}
            \textsc{Note.}\\ {Top 20 flare stars sorted by the number of flares on the corresponding star.
                The headers of this table are the same as in Table \ref{tab:starfrequency}.
            }\\
            \vspace{0.1em}
        \end{flushleft}

    \end{table*}

\begin{table*}[!h]
        \renewcommand{\arraystretch}{1}
        \addtolength{\tabcolsep}{+1pt}
        \caption{Properties of flare stars hosting planets.}
        \label{tab:planets}
        \centering
        \begin{tabular}{llcccl}
            \hline \hline
            \multicolumn{1}{c}{Host Star} & \multicolumn{1}{c}{Planet ID} & \multicolumn{1}{c}{Period $^a$} & Radius $^b$& Equilibrium Temp. $^c$ & \multicolumn{1}{c}{Information $^d$} \\
            \multicolumn{1}{c}{{\em TESS} ID} & \multicolumn{1}{c}{} & \multicolumn{1}{c}{(Days)} & ($R_\oplus$) & ($K$) & \multicolumn{1}{c}{} \\ \hline
            25078924 & TIC25078924.01 & 0.9097 & 14.43 & $-$ & Community {\em TESS} objects of interest \\
            25078924 & TIC25078924.02 & 0.9035 & 28.70 & $-$ & Community {\em TESS} objects of interest \\ \hline
            44797824 & TOI865.03 & 0.7456 $\pm$ 0.000018 & 3.79 $\pm$ 2.57 & 752 & {\em TESS} object of interest \\ \hline
            257605131 & TOI451.01 & 8.1855 $\pm$ 0.00177& 3.95 $\pm$ 0.69 & 640 & {\em TESS} object of interest \\
            257605131 & TIC257605131.02 & 1.8578 & 1.85 $\pm$ 0.06& 1337 & Community {\em TESS} objects of interest \\
            257605131 & TIC257605131.03 & 3.0643 & 1.82 $\pm$ 0.22& 1132 & Community {\em TESS} objects of interest \\  \hline
            373844472 & TOI275.01 & 0.9195 $\pm$ 0.000004& 17.94 $\pm$ 18.68
            & 1886 & {\em TESS} object of interest \\  \hline
            410214986$^*$ & DS Tuc A b & 8.1383 $\pm$ 0.000011& 5.70 $\pm$ 0.17& 850 & Confirmed planets \\ \hline
        \end{tabular}
        \begin{flushleft}
            \textsc{Note.}\\ {Planetary properties which are cross- matched with ExoFOP--{\em TESS}. Note that some values are not shown with errors as their error values are not included in the ExoFOP--{\em TESS} catalog.\\
                $^a$ Planet orbital period.\\
                $^b$ Radius of planet in units of Earth radius $R_{\oplus}$.\\
                $^c$ Equilibrium temperature, which represents the theoretical estimated temperature of planets heated by their host star.\\
                $^d$ Give information about each planet. Community {\em TESS} Objects of Interest (CTOIs) are planetary systems or potentially interesting targets identified by the community members, but not treated as a TESS Object of Interest (TOI) by the {\em TESS} project. TOI are {\em TESS} Objects of Interest, which are structured by the {\em TESS} Science Office (TSO) list, the Mikulski Archive for Space Telescopes (MAST) list and the ExoFOP--{\em TESS} list. \\
                $^*$Most of planets listed here are still not confirmed. DS Tuc A b is a hot planet, which is confirmed in the ExoFOP--{\em TESS} list \footnote{https://exoplanetarchive.ipac.caltech.edu/}. But the hosting star TIC410214986 is flagged as a binary star by {\em Hipparcos}.
            }\\
            \vspace{0.1em}
        \end{flushleft}

    \end{table*}

\begin{table*}[h]
    \renewcommand{\arraystretch}{1}
    \addtolength{\tabcolsep}{+0pt}
    \caption{Number fractions of Solar-type stars}
    \label{tab:N10result}
    \centering
    \begin{tabular}{ccclcc}
        \hline \hline
        \multirow{2}{*}{Data set}                       & \multicolumn{2}{c}{$5100 \mathrm{K} \leqslant T_{\mathrm{eff}}<5600 \mathrm{K}$}                            &  & \multicolumn{2}{c}{$5600 \mathrm{K} \leqslant T_{\mathrm{eff}}<6000 \mathrm{K}$}                            \\ \cline{2-3} \cline{5-6}
        & $N_{\rm star}(P < 10\; { \rm days})/N_{\rm all}$ & $N_{\rm star}(P \geqslant 10\; { \rm days})/N_{\rm all}$ &  & $N_{\rm star}(P < 10\; { \rm days})/N_{\rm all}$ & $N_{\rm star}(P \geqslant 10\; { \rm days})/N_{\rm all}$ \\ \hline
        Gyrochronology relation$^a$                     & $5.1\%-7.1\%$                                    & $92.9\%-94.9\%$                                          &  & $7.1\%-10.9\%$                                   & $89.1\%-92.9\%$                                          \\
        \cite{2019ApJ...876...58N}$^b$ & $14.1\%$                                         & $85.9\%$                                                 &  & $21.7\%$                                         & $78.3\%$                                                 \\
        This work                                       & $82.6\%$                                         & $17.4\%$                                                 &  & $87.6\%$                                         & $12.4\%$                                                 \\ \hline
        \multicolumn{1}{l}{}                            & \multicolumn{1}{l}{}                             & \multicolumn{1}{l}{}                                     &  & \multicolumn{1}{l}{}                             & \multicolumn{1}{l}{}
    \end{tabular}
    \begin{flushleft}
        \textsc{Note.}\\ {Here, we separate solar-type stars into two parts according to their surface effective temperatures. $N_{\rm star}(P < 10\; { \rm days})/N_{\rm all}$ denotes the number fraction between stars with period less than 10 days and all stars. (a) gives the results calculated by Equation \ref{equ:NpNall}. (b) lists the results shown in Table 9 of \citet{2019ApJ...876...58N}, who used the reported stellar periods in \citet{2014ApJS..211...24M}.
        }\\
        \vspace{0.1em}
    \end{flushleft}

\end{table*}


\end{document}